%% file: conf-entcs.tex
\documentclass[copyright,creativecommons]{eptcs} 
\providecommand{\event}{QAPL 2016}

\usepackage{times}
\usepackage{mathptmx}
\usepackage{etoolbox}
\usepackage{amstext}
\usepackage{amsmath}
\usepackage{figbox}
\usepackage{figspace}
\usepackage{supertabular}
\usepackage{pifont}
\usepackage{latexsym}
\usepackage{amsfonts}
\usepackage{extarrows}
\usepackage{graphicx}
\usepackage{wrapfig}
\usepackage{breakurl}

\usepackage{subfigure}
\usepackage{amssymb}

\usepackage{dsfont}

\usepackage{placeins}
\usepackage{comment}
\usepackage{proof}

\usepackage{shadowtext}

\input defns

\input catmac

\input mathdefs

\newtoggle{short}
\toggletrue{short}

\def\titlerunning{Parameterized Dataflow}
\def\authorrunning{D.~Duggan \&\ J.~Yao}

\title{Parameterized Dataflow \\ {\small Extended Abstract}}
\author{Dominic Duggan
\institute{
Department of Computer Science, \\
Stevens Institute of Technology \\
Hoboken, New Jersey, USA. }
\email{dduggan@stevens.edu}
\and
Jianhua Yao
\institute{
Department of Computer Science, \\
Stevens Institute of Technology \\
Hoboken, New Jersey, USA. }
\email{jyao1@stevens.edu}
}

\begin{document}

\maketitle

\input abstract

\input intro

\input dataflow

\input related

\input concl

\bibliographystyle{eptcs}
\bibliography{biblio}

\pagebreak

\appendix

\end{document}

%% file: defns.tex
\newcommand{\inflab}[1]{\text{\sc #1}}

\newcommand{\Mid}{\mbox{$\hspace{0.1in}\mid\hspace{0.1in}$}}

\newcommand{\mysize}[1]{\text{$\mid #1 \mid$}}
\renewcommand{\emptyset}{\ensuremath{\{\}}}
\newcommand{\subst}[3]{\ensuremath{\{#2/#3\}#1}}

\newcommand{\myiff}{\ensuremath{\Longleftrightarrow}}

\newcommand{\Hskip}{\hspace{0.15in}}
\newcommand{\Vskip}{\vspace{0.15in}}

\newcommand{\Sdata}{\ensuremath{\mathbf{S}_{\mathbf{DATA}}}}

\newcommand{\kw}[1]{\ensuremath{\mathsf{#1}}}
\newcommand{\kwK}[1]{\ensuremath{\mathsf{\text{\shadowtext{#1}}}}}
\newcommand{\seq}[1]{\ensuremath{\overrightarrow{#1}}}

\newcommand{\m}{\tt}
\newcommand{\makecode}[1]{\text{\m #1}}        
\newcommand{\mc}[1]{\makecode{#1}}             

        {\end{tabbing}\end{trivlist}}              
        {\end{tabbing}\end{trivlist}}              
        {\end{tabbing}\end{trivlist}}              

\newcommand{\numberTy}[1]{\ensuremath{\underline{#1}}}
\newcommand{\inftyTy}{\ensuremath{\numberTy{\infty}}}

\newcommand{\chanLab}{\ensuremath{c}}

\newcommand{\parsym}{\ensuremath{\parallel}}

\newcommand{\recv}[1]{\ensuremath{#1?}}
\newcommand{\send}[2]{\ensuremath{#1!#2}}

\newcommand{\stopProc}{\ensuremath{\kw{stop}}}

\newcommand{\tyenv}{\ensuremath{\Gamma}}
\newcommand{\valenv}{\ensuremath{\Delta}}

\newcommand{\STy}{\ensuremath{\tau}}
\newcommand{\Ty}{\ensuremath{\STy}}
\newcommand{\Kind}{\ensuremath{\kappa}}

\newcommand{\NetworkSym}{\ensuremath{\mathit{N}}}
\newcommand{\NetworkSimple}[4]{\ensuremath{\kw{network}(#1,#2\vdash #3:#4)}}
\newcommand{\NetworkSimpleSym}{\NetworkSym}
\newcommand{\NetworkSimpleSig}[3]{\ensuremath{\kw{Network}(#1,#2\rhd
    #3)}}
\newcommand{\NetworkSimpleSigSym}{\ensuremath{\mathit{NS}}}

\newcommand{\CausalitySetSym}{\ensuremath{\Kappa}}

\newcommand{\boolTy}{\ensuremath{\kw{Boolean}}}

\newcommand{\intTy}{\ensuremath{\kw{Integer}}}

\newcommand{\refTy}[1]{\ensuremath{\kw{Ref}(#1)}}

\newcommand{\typeK}{\ensuremath{\kwK{Type}}}
\newcommand{\channelK}{\ensuremath{\kwK{Channel}(\channelDelayFlag,\STy_{\mathit{limit}})}}
\newcommand{\channelKWithDelay}[1]{\ensuremath{\kwK{Channel}({#1,\STy_{\mathit{limit}}})}}

\newcommand{\channelKWithDelayLimit}[2]{\ensuremath{\kwK{Channel}({#1,#2})}}
\newcommand{\channelTy}[3]{\ensuremath{\kw{Channel}({#1,#2,#3})}}

\newcommand{\polarity}{\ensuremath{\pi}}
\newcommand{\posPolarity}{\mc{+}}
\newcommand{\negPolarity}{\mc{-}}
\newcommand{\pmPolarity}{\ensuremath{\pm}}

\newcommand{\posChan}[1]{\ensuremath{#1}}
\newcommand{\negChan}[1]{\ensuremath{#1}}

\newcommand{\channelDelayFlag}{\ensuremath{b}}
\newcommand{\channelDelay}{\ensuremath{1}}
\newcommand{\channelNoDelay}{\ensuremath{0}}

\newcommand{\flowstateK}{\ensuremath{\kwK{Flowstate}}}

\newcommand{\Array}{\ensuremath{A}}
\newcommand{\arrayVal}[1]{\ensuremath{\langle #1\rangle}}
\newcommand{\assignArrayVal}[3]{\ensuremath{#1[#2\mapsto #3]}}

\newcommand{\FlowstateSym}{\ensuremath{FS}}
\newcommand{\ActorFlowstateSym}{\ensuremath{\mathit{AS}}}
\newcommand{\ProcFlowstateSym}{\ensuremath{\FlowstateSym}}

\newcommand{\emptyFlowstate}{\ensuremath{\varepsilon}}
\newcommand{\seqFlowstate}[2]{\ensuremath{#1\seqAS #2}}
\newcommand{\parFlowstate}[2]{\ensuremath{#1\parsym #2}}

\newcommand{\seqAS}{\ensuremath{;}}

\newcommand{\tyenvj}[1]{\ensuremath{\vdash #1\ \kw{ok}}}
\newcommand{\valenvj}[2]{\ensuremath{#1\vdash #2\ \kw{ok}}}
\newcommand{\kindj}[2]{\ensuremath{#1\vdash #2\ \kw{ok}}}
\newcommand{\tyj}[3]{\ensuremath{#1\vdash #2: #3}}
\newcommand{\sizeleqj}[3]{\ensuremath{#1 \vdash #2\leq #3}}
\newcommand{\eventj}[2]{\ensuremath{#1\vdash #2\ \kw{ok}}}

\newcommand{\flowj}[2]{\ensuremath{#1\vdash #2:\flowstateK}}
\newcommand{\iterj}[2]{\ensuremath{#1\vdash #2\ \kw{ok}}}
\newcommand{\guardj}[2]{\ensuremath{#1\vdash #2\ \kw{ok}}}
\newcommand{\valj}[4]{\ensuremath{#1;#2\vdash #3:#4}}
\newcommand{\expj}[5]{\ensuremath{#1;#2\vdash #3:#4{::}#5}}
\newcommand{\procj}[4]{\ensuremath{#1;#2\vdash #3:#4}}

\newcommand{\iredProc}[1]{\ensuremath{\xlongrightarrow{#1}}}
\newcommand{\iredExp}{\ensuremath{\longrightarrow}}

\newcommand{\redProc}[2]{\ensuremath{\xlongrightarrow{#1,#2}}}
\newcommand{\redExp}[2]{\ensuremath{\xlongrightarrow{#1,#2}}}

\newcommand{\optRedProc}[2]{\ensuremath{\xlongrightarrow{[#1,#2]}}}
\newcommand{\optRedExp}[2]{\ensuremath{\xlongrightarrow{[#1,#2]}}}

\newcommand{\eventSym}{\ensuremath{\alpha}}
\newcommand{\sendEvent}[1]{\ensuremath{#1!}}
\newcommand{\recvEvent}[1]{\ensuremath{#1?}}

\newcommand{\Proc}{\ensuremath{P}}
\newcommand{\Exp}{\ensuremath{E}}
\newcommand{\val}{\ensuremath{V}}

\newcommand{\newRef}[1]{\ensuremath{\kw{ref}(#1)}}
\newcommand{\derefExp}[1]{\ensuremath{* #1}}
\newcommand{\assignExp}[2]{\ensuremath{#1 := #2}}

\newcommand{\ProcCtxt}{\ensuremath{\mathbb{P}}}
\newcommand{\ExpCtxt}{\ensuremath{\mathbb{E}}}

\newcommand{\Heap}{\ensuremath{H}}
\newcommand{\HeapTy}{\ensuremath{\mathit{HT}}}
\newcommand{\emptyHeap}{\varepsilon}
\newcommand{\heapBind}[2]{\ensuremath{#1\mapsto #2}}
\newcommand{\heapJoin}[2]{\ensuremath{#1\uplus #2}}

\newcommand{\heapTyBind}[2]{\ensuremath{#1: #2}}
\newcommand{\heapTyJoin}[2]{\ensuremath{#1\uplus #2}}

\newcommand{\Buffer}{\ensuremath{B}}
\newcommand{\BufferTy}[2]{#2[#1]}
\newcommand{\emptyBuff}[1]{\ensuremath{\varepsilon_{#1}}}
\newcommand{\buffCell}[2]{{[#2]\ensuremath{_{#1}}}}
\newcommand{\joinBuff}[3]{{#2 \ensuremath{ @_{#1}} #3}}
\newcommand{\consBuff}[3]{{#2 \ensuremath{::_{#1}} #3}}

\newcommand{\sizeBuff}[1]{\ensuremath{\mysize{#1}}}
\newcommand{\maxsizeBuff}[1]{\ensuremath{\mathit{size}(#1)}}

\newcommand{\emptyEnv}{\ensuremath{\varepsilon}}

\newcommand{\heapj}[4]{\ensuremath{#1;#2\vdash #3:#4}}
\newcommand{\buffj}[4]{\ensuremath{#1;#2\vdash #3:#4}}

\newcommand{\sizeK}[1]{\ensuremath{\kwK{Size}(#1)}}

\newcommand{\channelArrayK}[1]{\ensuremath{\kwK{Channel}(\channelDelayFlag,\STy_{\mathit{limit}})[#1]}}

\newcommand{\channelArrayKWithDelayLimit}[3]{\ensuremath{\kwK{Channel}({#1,#2})[#3]}}

\newcommand{\channelArrayTy}[4]{\ensuremath{\kw{Channel}(#1,#2,#3)[#4]}}

\newcommand{\sizeTy}[1]{\ensuremath{\kw{Size}(#1)}}
\newcommand{\sizeExp}[1]{\ensuremath{\kw{size}(#1)}}
\newcommand{\fromSizeExp}[1]{\ensuremath{\kw{fromSize}(#1)}}
\newcommand{\indexTy}[1]{\ensuremath{\kw{Index}(#1)}}
\newcommand{\indexExp}[1]{\ensuremath{\kw{index}(#1)}}
\newcommand{\fromIndexExp}[1]{\ensuremath{\kw{fromIndex}(#1)}}

\newcommand{\comprehension}[2]{\ensuremath{\{#1\mid #2\}}}
\newcommand{\eventComp}[2]{\comprehension{#1}{#2}}
\newcommand{\actorComp}[5]{\comprehension{#1}{#2,#3\leftarrow
    #4{..}#5}}
\newcommand{\fsComp}[4]{\comprehension{#1}{#2\leftarrow
    #3{..}#4}}

\newcommand{\eventArity}[2]{\ensuremath{#1\ast #2}}

\newcommand{\lambdar}[2]{\ensuremath{#1\Rightarrow}}
\newcommand{\funtyopr}[2]{\ensuremath{\xrightarrow{#1}}}
\newcommand{\actorpcr}[2]{\ensuremath{#1}}

\newcommand{\iterType}[3]{\ensuremath{#1\leftarrow #2..#3}}
\newcommand{\iterTypeSym}{\ensuremath{\mathcal{I}}}
\newcommand{\guardTypeSym}{\ensuremath{\mathcal{G}}}
\newcommand{\relop}{\ensuremath{\rho}}
\newcommand{\divides}{\ensuremath{\mid}}
\newcommand{\wfguard}[3]{\ensuremath{\mathit{wfguard}(#1#2#3)}}

\newcommand{\livefsj}[5]{\ensuremath{#1\vdash (#2,#3)\longrightarrow
    (#4,#5)}}
\newcommand{\detfsj}[2]{\ensuremath{#1\vdash #2\
    \mathbf{deterministic}}}
\newcommand{\eventCompl}[1]{\ensuremath{\overline{#1}}}

\newcommand{\Config}[2]{\ensuremath{(#1,#2)}}

\newcommand{\ConfigSym}{\ensuremath{C_{\bf DATA}}}

\newcommand{\aredTyj}[5]{\ensuremath{#3 \xlongrightarrow{#4} #5}}
\newcommand{\airedTyj}[4]{\ensuremath{#3 \longrightarrow #4}}

%% file: catmac.tex
\newcount\yyitercn 
\newcount\yycnA 

\newbox\yybxA 
\newbox\yybxB 
\newbox\yybxC 
\newbox\yybxD 
\newbox\yybxE 
\newbox\yybxF 
\newbox\yybxG 

\newdimen\yydmA 
\newdimen\yydmB 
\newdimen\yydmC 
\newdimen\yydmD 
\newdimen\yydmE 
\newdimen\yydmF 
\newdimen\yydmG 
\newdimen\yydmH 
\newdimen\yydmI 

\def\defaxis#1#2{\setbox\yybxA=\hbox{{$#2\vcenter{}$}}
\edef#1{\the\ht\yybxA}}

\defaxis\displayaxis\displaystyle
\defaxis\textaxis\textstyle
\defaxis\scriptaxis\scriptstyle
\defaxis\scriptscriptaxis\scriptscriptstyle

\def\godmult{\mathbin{\mathchoice
{\kern0.25pt\vrule height7pt depth3.5pt
\vrule height7pt depth-6.6pt width4pt
\kern-4pt\vrule height-3.1pt depth3.5pt width4pt
\kern-3.3pt\textstyle ;\kern-0.03333pt
\vrule height7pt depth3.5pt\kern1pt}
{\kern0.25pt\vrule height7pt depth3.5pt
\vrule height7pt depth-6.6pt width4pt
\kern-4pt\vrule height-3.1pt depth3.5pt width4pt
\kern-3.3pt\textstyle ;\kern-0.03333pt
\vrule height7pt depth3.5pt\kern1pt}
{\kern0.5pt\vrule height5pt depth2.5pt
\vrule height5pt depth-4.6pt width3pt
\kern-3pt\vrule height-2.1pt depth2.5pt width3pt
\kern-2.4pt\scriptstyle ;\kern0.03889pt
\vrule height5pt depth2.5pt\kern1pt}
{\kern0.5pt\vrule height4.25pt depth2pt
\vrule height4.25pt depth-3.85pt width2.75pt
\kern-2.75pt\vrule height-1.6pt depth2pt width2.75pt
\kern-2.275pt\scriptscriptstyle ;\kern-0.00277pt
\vrule height4.25pt depth2pt\kern1pt}}}

\def\yyarrowb#1#2#3#4#5#6#7#8#9{{ 
\setbox\yybxA=\hbox{$\mathsurround=0pt #2
\mkern #7\mathop{}\limits #4\mkern #8$}
\yydmA=\wd\yybxA\relax
\setbox\yybxA=\hbox{$\mathsurround=0pt #2 #9$}
\ifdim\yydmA<\wd\yybxA\relax\yydmA=\wd\yybxA\relax\else\fi
\setbox\yybxA=\hbox to \yydmA{#1 #2}
\yydmA=#3\relax\advance\yydmA by #5\relax\ht\yybxA=\yydmA\relax
\yydmA=-#3\relax\advance\yydmA by #6\relax\dp\yybxA=\yydmA\relax
#2\mathop{\box\yybxA}\limits #4}}

\def\defbig#1#2#3#4{\setbox\yybxA=\hbox{$#1#2($}
\yydmA=\ht\yybxA\relax\yydmB=\dp\yybxA\relax\yycnA=50\relax
\yydefbiga{#1}{#3}{#4}}

\def\yydefbiga#1#2#3{\advance\yydmA by 0.2pt\relax 
\advance\yycnA by -1\relax
\setbox\yybxA=\hbox{$#1\left(\vbox to \yydmA{}\right.
\nulldelimiterspace=0pt\mathsurround=0pt$}
\yydmC=\dp\yybxA\relax\advance\yydmC by -\yydmB\relax
\advance\yydmC by -#3\relax
\ifdim\yydmC<0pt\relax\ifnum\yycnA>0\relax
\yydefbiga{#1}{#2}{#3}
\else\yydefbigb{#2}{#1}\fi
\else\yydefbigb{#2}{#1}\fi}

\def\yydefbigb#1#2{\edef\yybigsize{{\the\yydmA}} 
\expandafter\yydefbigc\yybigsize{#1}{#2}}

\def\yydefbigc#1#2#3{\def#2##1{{\hbox{$#3\left##1\vbox to #1{}\right. 
\nulldelimiterspace=0pt\mathsurround=0pt$}}}}

\defbig\textstyle\relax\big{0.95pt}
\defbig\textstyle\big\Big{0.95pt}
\defbig\textstyle\Big\bigg{0.95pt}
\defbig\textstyle\bigg\Bigg{0.95pt}

\defbig\scriptstyle\relax\scriptbig{0.95pt}
\defbig\scriptscriptstyle\relax\scriptscriptbig{0.95pt}

\def\normsize#1{\begingroup\let\currbig=\relax #1\endgroup}
\def\bigsize#1{\begingroup\let\currbig=\big #1\endgroup}
\def\Bigsize#1{\mathinner{{\let\currbig=\Big #1}}}
\def\biggsize#1{\mathinner{{\let\currbig=\bigg #1}}}
\def\Biggsize#1{\mathinner{{\let\currbig=\Bigg #1}}}
\def\scriptbigsize#1{\begingroup\let\currbig=\scriptbig #1\endgroup}
\def\scriptscriptbigsize#1{\begingroup\let\currbig=\scriptscriptbig
#1\endgroup}

\let\currbig=\relax

\def\varlbrace{\mathopen\currbig\lbrace}
\def\varrbrace{\mathclose\currbig\rbrace}

\def\yykernparta#1#2{ 
\setbox\yybxA=\hbox{$\mathsurround=0pt\relax #1$}
\yydmA=\wd\yybxA\multiply\yydmA by #2\relax
\divide\yydmA by 100\relax \kern\yydmA}

\def\yyemboldena#1#2{ 
\setbox\yybxA=\hbox{$\mathsurround=0pt\relax #1$}
\copy\yybxA\kern-\wd\yybxA\kern #2\relax
\raise #2\copy\yybxA\kern-\wd\yybxA\relax
\raise -#2\copy\yybxA\kern-\wd\yybxA\kern #2\relax
\box\yybxA}

\def\simplist#1#2#3#4#5{
\yyitercn=#1\relax
\def\yyrightdel{#3}
\def\yysepsym{#4}
\ifnum\yyitercn=0\relax
#5 \let\next=\relax
\else #2 \let\next=\yysimplista
\fi\next}

\def\yysimplista#1{\begingroup\normsize{#1}\endgroup 
\advance\yyitercn by -1\relax
\ifnum\yyitercn=0\relax
\yyrightdel \let\next=\relax
\else \yysepsym \let\next=\yysimplista
\fi\next}

\def\emptyset{\varlbrace\varrbrace}

\def\pairlist#1#2#3#4#5#6{
\yyitercn=#1\relax
\def\yyrightdel{#3}
\def\yysepsym{#4}
\def\yypairsym{#5}
\ifnum\yyitercn=0\relax
#6 \let\next=\relax
\else #2 \let\next=\yypairlista
\fi\next}

\def\yypairlista#1#2{ 
\begingroup\normsize{#1}\endgroup\yypairsym
\begingroup\normsize{#2}\endgroup
\advance\yyitercn by -1\relax
\ifnum\yyitercn=0\relax
\yyrightdel \let\next=\relax
\else \yysepsym \let\next=\yypairlista
\fi\next}

\def\infix#1#2{
\yyitercn=#1\relax
\def\yysepsym{#2}
\ifnum\yyitercn=0\relax
\errmessage{CALL OF \string\infix0}
\let\next=\relax
\else\let\next=\yyinfixa
\fi\next}

\def\yyinfixa#1{\begingroup #1\endgroup 
\advance\yyitercn by -1\relax
\ifnum\yyitercn=0\relax
\let\next=\relax
\else \yysepsym \let\next=\yyinfixa
\fi\next}

\def\expandlargeops{\def\yylopstyle{\displaystyle}
\def\yylopstyleswitch{1}}

\expandlargeops

\let\oldprod=\prod
\let\oldsum=\sum

\def\Union{\mathop{\mathchoice
{\yylopstyle\bigcup}
{\textstyle\bigcup}
{\raise-0.19355ex\hbox{$\mathsurround=0pt{\textstyle\cup}$}}
{\raise-0.13548ex\hbox{$\mathsurround=0pt{\scriptstyle\cup}$}}}}

\def\Intersect{\mathop{\mathchoice
{\yylopstyle\bigcap}
{\textstyle\bigcap}
{\raise-0.19355ex\hbox{$\mathsurround=0pt{\textstyle\cap}$}}
{\raise-0.13548ex\hbox{$\mathsurround=0pt{\scriptstyle\cap}$}}}}

\def\prod{\mathop{\mathchoice
{\yylopstyle\oldprod}
{\textstyle\oldprod}
{\scriptstyle\Pi}
{\scriptscriptstyle\Pi}}}

\def\sum{\mathop{\mathchoice
{\yylopstyle\oldsum}
{\textstyle\oldsum}
{\scriptstyle\Sigma}
{\scriptscriptstyle\Sigma}}}

\def\Lub{\mathop{\mathchoice
{\yylopstyle\bigsqcup}
{\textstyle\bigsqcup}
{\raise-0.19355ex\hbox{$\mathsurround=0pt{\textstyle\sqcup}$}}
{\raise-0.13548ex\hbox{$\mathsurround=0pt{\scriptstyle\sqcup}$}}}}

\def\Glb{\mathop{\mathchoice
{\ifnum\yylopstyleswitch=1\relax
{\kern0.25pt
\vrule height10.53pt depth3.25pt width0.7pt
\vrule height10.53pt depth-9.83pt width8.25pt
\vrule height10.53pt depth3.25pt width0.7pt
\vrule height11.0pt depth5.0pt width0pt\kern1.2111pt}
\else {\setbox\yybxB=\hbox{\kern0.25pt
\vrule height8.37pt depth1.2pt width0.7pt
\vrule height8.37pt depth-7.67pt width5.5pt
\vrule height8.37pt depth1.2pt width0.7pt\kern1.18333pt}
\ht\yybxB=8.0pt\dp\yybxB=2.0pt\relax\box\yybxB}
\fi}
{{\setbox\yybxB=\hbox{\kern0.25pt
\vrule height8.37pt depth1.2pt width0.7pt
\vrule height8.37pt depth-7.67pt width5.5pt
\vrule height8.37pt depth1.2pt width0.7pt\kern1.18333pt}
\ht\yybxB=8.0pt\dp\yybxB=2.0pt\relax\box\yybxB}}
{{\raise-1pt\hbox{$\mathsurround=0pt{\textstyle\sqcap}$}}}
{{\raise-0.7pt\hbox{$\mathsurround=0pt{\scriptstyle\sqcap}$}}}}}

\def\multscript#1#2#3#4#5#6#7#8#9{{
\setbox\yybxA=\hbox{$\mathsurround=0pt\mathop{{#1}}$}
\setbox\yybxB=\hbox{$\mathsurround=0pt#9{{#2}}$}
\setbox\yybxC=\hbox{$\mathsurround=0pt#9{{#3}}$}
\setbox\yybxD=\hbox{$\mathsurround=0pt#9{{#4}}$}
\setbox\yybxE=\hbox{$\mathsurround=0pt#9{{#5}}$}
\yydmA=\ht\yybxA
\yydmB=\dp\yybxA
\yydmC=\yydmA\advance\yydmC by #6
\yydmD=\yydmB\advance\yydmD by #6
\yydmE=\wd\yybxA
\yydmF=\yydmE\advance\yydmF by 0pt
\ht\yybxA=\yydmC\dp\yybxA=\yydmD
\setbox\yybxF=\hbox{$\mathsurround=0pt#8\mathop{\box\yybxA}\limits
\ifdim\wd\yybxB=0pt\else\ifdim\wd\yybxB>\yydmF\else _{{#2}}\fi\fi
\ifdim\wd\yybxC=0pt\else\ifdim\wd\yybxC>\yydmF\else ^{{#3}}\fi\fi$}%
\yydmG=\ht\yybxF
\yydmH=\dp\yybxF
\yydmI=\wd\yybxF
\ht\yybxF=\yydmA\dp\yybxF=\yydmB
\setbox\yybxG=\hbox{$\mathsurround=0pt#8\mathop{\box\yybxF}\nolimits
\ifdim\wd\yybxD=0pt\else _{{#4}}\fi
\ifdim\wd\yybxE=0pt\else ^{{#5}}\fi
\vrule height\yydmC depth\yydmD width0pt$}%
\yydmA=\wd\yybxG
\yydmB=\yydmE\advance\yydmB by -\yydmI\divide\yydmB by 2\relax
\advance\yydmB by #7\relax\ifdim\yydmB<0pt\yydmB=0pt\fi
\yydmC=\yydmB \advance \yydmC by \yydmI \advance\yydmC by -\yydmA
\setbox\yybxD=\hbox{$\mathsurround=0pt#8\mathop
{\kern\yydmB\box\yybxG\kern\yydmC}\limits
\ifdim\wd\yybxB=0pt\else\ifdim\wd\yybxB>\yydmF _{{#2}}\fi\fi
\ifdim\wd\yybxC=0pt\else\ifdim\wd\yybxC>\yydmF ^{{#3}}\fi\fi$}%
\yydmD=\yydmI\advance\yydmD by \wd\yybxD\divide\yydmD by -2\relax
\advance\yydmD by \yydmA\relax\ifdim\yydmD<0pt\yydmD=0pt\fi
\box\yybxD\kern\yydmD\vrule height\yydmG depth\yydmH width0pt}}

\def\compscript#1#2#3#4#5#6#7#8#9{{
\setbox\yybxA=\hbox{$\mathsurround=0pt\mathop{{#1}}$}
\setbox\yybxB=\hbox{$\mathsurround=0pt#9{{#2}}$}
\setbox\yybxC=\hbox{$\mathsurround=0pt#9{{#3}}$}
\setbox\yybxD=\hbox{$\mathsurround=0pt#9{{#4}}$}
\setbox\yybxE=\hbox{$\mathsurround=0pt#9{{#5}}$}
\yydmA=\ht\yybxA\advance\yydmA by #6
\yydmB=\dp\yybxA\advance\yydmB by #6
\yydmC=\wd\yybxA
\yydmD=\wd\yybxE\advance\yydmD by -\wd\yybxD
\setbox\yybxF=\hbox{$\mathsurround=0pt#8\mathop{\box\yybxA}\nolimits
\ifdim\wd\yybxD=0pt\else _{{#4}}\fi
\ifdim\wd\yybxE=0pt\else ^{{#5}}\fi$}%
\yydmE=\wd\yybxF\relax
\ifdim\wd\yybxB=0pt\ifdim\wd\yybxC=0pt
\setbox\yybxG=\box\yybxF%
\else
\setbox\yybxG=\hbox{$\mathsurround=0pt#8\mathop{\box\yybxF
\ifdim\yydmD<0pt\kern\yydmD\else\fi}\nolimits ^{{#3}}$}%
\fi\else
\ifdim\wd\yybxD=0pt\else\yydmF=\dp\yybxF
\advance\yydmF by 1pt\dp\yybxF=\yydmF\relax\fi
\ifdim\wd\yybxC=0pt
\setbox\yybxG=\hbox{$\mathsurround=0pt#8\mathop{\box\yybxF
\ifdim\yydmD>0pt\kern-\yydmD\else\fi}\nolimits _{{#2}}$}%
\else
\setbox\yybxG=\hbox{$\mathsurround=0pt#8\mathop{\box\yybxF
}\nolimits _{{#2}}^{{#3}}$}%
\fi\fi
\advance\yydmE by -\wd\yybxG%
\advance\yydmC by -\wd\yybxG\advance\yydmC by #7\relax%
\ifdim\yydmE<0pt\yydmE=0pt\else\fi\relax%
\ifdim\yydmC>\yydmE\yydmE=\yydmC\else\fi
\kern #7\box\yybxG\kern\yydmE\vrule height\yydmA depth\yydmB width0pt}}

\def\Lubms#1#2#3#4{\mathop{\mathchoice
{\multscript{\yylopstyle\bigsqcup}{#1}{#2}{#3}{#4}
{0pt}{0pt}{\yylopstyle}{\scriptstyle}}
{\compscript{\textstyle\bigsqcup}{#1}{#2}{#3}{#4}
{0pt}{0pt}{\textstyle}{\scriptstyle}}
{\compscript
{\raise-0.19355ex\hbox{$\mathsurround=0pt{\textstyle\sqcup}$}}
{#1}{#2}{#3}{#4}{0pt}{0pt}{\scriptstyle}{\scriptscriptstyle}}
{\compscript
{\raise-0.13548ex\hbox{$\mathsurround=0pt{\scriptstyle\sqcup}$}}
{#1}{#2}{#3}{#4}{0pt}{0pt}{\scriptscriptstyle}{\scriptscriptstyle}}}}

\def\Glbms#1#2#3#4{\mathop{\mathchoice
{\ifnum\yylopstyleswitch=1\relax
\multscript{\kern0.25pt
\vrule height10.53pt depth3.25pt width0.7pt
\vrule height10.53pt depth-9.83pt width8.25pt
\vrule height10.53pt depth3.25pt width0.7pt
\vrule height11.0pt depth5.0pt width0pt\kern1.2111pt}
{#1}{#2}{#3}{#4}{0pt}{0pt}{\displaystyle}{\scriptstyle}
\else \multscript{\setbox\yybxB=\hbox{\kern0.25pt
\vrule height8.37pt depth1.2pt width0.7pt
\vrule height8.37pt depth-7.67pt width5.5pt
\vrule height8.37pt depth1.2pt width0.7pt\kern1.18333pt}
\ht\yybxB=8.0pt\dp\yybxB=2.0pt\relax\box\yybxB}
{#1}{#2}{#3}{#4}{0pt}{0pt}{\textstyle}{\scriptstyle}\fi}
{\compscript{\setbox\yybxB=\hbox{\kern0.25pt
\vrule height8.37pt depth1.2pt width0.7pt
\vrule height8.37pt depth-7.67pt width5.5pt
\vrule height8.37pt depth1.2pt width0.7pt\kern1.18333pt}
\ht\yybxB=8.0pt\dp\yybxB=2.0pt\relax\box\yybxB}
{#1}{#2}{#3}{#4}{0pt}{0pt}{\textstyle}{\scriptstyle}}
{\compscript
{\raise-1pt\hbox{$\mathsurround=0pt{\textstyle\sqcap}$}}
{#1}{#2}{#3}{#4}{0pt}{0pt}{\scriptstyle}{\scriptscriptstyle}}
{\compscript
{\raise-0.7pt\hbox{$\mathsurround=0pt{\scriptstyle\sqcap}$}}
{#1}{#2}{#3}{#4}{0pt}{0pt}{\scriptscriptstyle}{\scriptscriptstyle}}}}

\def\encircle#1#2#3#4{\setlength{\unitlength}{0.1pt}
\begin{picture}(0,0)
\put(#1,#2){\circle{#3}}
\end{picture}
{#4}}

\def\cirprodmsnorm#1#2#3#4{\mathop{\mathchoice
{\ifnum\yylopstyleswitch=1\relax
\multscript
{\kern3.5pt\encircle{59}{34}{200}{\displaystyle\oldprod}\kern3.5pt}
{#1}{#2}{#3}{#4}{2.5pt}{0.5pt}{\displaystyle}{\scriptstyle}
\else \multscript
{\kern2.5pt\encircle{42}{34}{140}{\textstyle\oldprod}\kern2.5pt}
{#1}{#2}{#3}{#4}{2.0pt}{0pt}{\textstyle}{\scriptstyle}\fi}
{\compscript
{\kern2.5pt\encircle{42}{34}{140}{\textstyle\oldprod}\kern2.5pt}
{#1}{#2}{#3}{#4}{1.5pt}{0pt}{\textstyle}{\scriptstyle}}
{\compscript
{\kern1.75pt\encircle{31}{27}{90}{\scriptstyle\Pi}\kern1.75pt}
{#1}{#2}{#3}{#4}{1pt}{0.5pt}{\scriptstyle}{\scriptscriptstyle}}
{\compscript
{\kern1pt\encircle{23}{20}{70}{\scriptscriptstyle\Pi}\kern1pt}
{#1}{#2}{#3}{#4}{0.5pt}{0.25pt}{\scriptscriptstyle}{\scriptscriptstyle}}}}

\def\cirsummsnorm#1#2#3#4{\mathop{\mathchoice
{\ifnum\yylopstyleswitch=1\relax
\multscript
{\kern2.5pt\encircle{65}{33}{200}{\displaystyle\oldsum}\kern2.5pt}
{#1}{#2}{#3}{#4}{2.5pt}{1.0pt}{\displaystyle}{\scriptstyle}
\else \multscript
{\kern2.5pt\encircle{47}{34}{150}{\textstyle\oldsum}\kern2.5pt}
{#1}{#2}{#3}{#4}{2.75pt}{0pt}{\textstyle}{\scriptstyle}\fi}
{\compscript
{\kern2.5pt\encircle{47}{34}{150}{\textstyle\oldsum}\kern2.5pt}
{#1}{#2}{#3}{#4}{2.25pt}{0pt}{\textstyle}{\scriptstyle}}
{\compscript
{\kern1.25pt\encircle{27}{28}{80}{\scriptstyle\Sigma}\kern1.25pt}
{#1}{#2}{#3}{#4}{1pt}{0.5pt}{\scriptstyle}{\scriptscriptstyle}}
{\compscript
{\kern1.25pt\encircle{25}{19}{70}{\scriptscriptstyle\Sigma}\kern1.25pt}
{#1}{#2}{#3}{#4}{0.5pt}{0.5pt}{\scriptscriptstyle}{\scriptscriptstyle}}}}

\def\cirLmmsnorm#1#2#3#4{\mathop{\mathchoice
{\multscript
{\kern2pt\encircle{87}{37}{200}{\textstyle{\rm Lm}}\kern2pt}
{#1}{#2}{#3}{#4}{6pt}{0pt}{\textstyle}{\scriptstyle}}
{\compscript
{\kern2pt\encircle{87}{37}{200}{\textstyle{\rm Lm}}\kern2pt}
{#1}{#2}{#3}{#4}{6pt}{0pt}{\textstyle}{\scriptstyle}}
{\compscript
{\kern1.5pt\encircle{61}{27}{140}{\scriptstyle{\rm Lm}}\kern1.5pt}
{#1}{#2}{#3}{#4}{3.5pt}{0.5pt}{\scriptstyle}{\scriptscriptstyle}}
{\compscript
{\kern1pt\encircle{51}{20}{120}{\scriptscriptstyle{\rm Lm}}\kern1pt}
{#1}{#2}{#3}{#4}{3pt}{0.3pt}{\scriptscriptstyle}{\scriptscriptstyle}}}}

\def\cirColmsnorm#1#2#3#4{\mathop{\mathchoice
{\multscript
{\kern2pt\encircle{93}{35}{200}{\textstyle{\rm Col}}\kern2pt}
{#1}{#2}{#3}{#4}{6.25pt}{0pt}{\textstyle}{\scriptstyle}}
{\compscript
{\kern2pt\encircle{93}{35}{200}{\textstyle{\rm Col}}\kern2pt}
{#1}{#2}{#3}{#4}{6.25pt}{0pt}{\textstyle}{\scriptstyle}}
{\compscript
{\kern1.5pt\encircle{65}{24}{140}{\scriptstyle{\rm Col}}\kern1.5pt}
{#1}{#2}{#3}{#4}{4pt}{0.5pt}{\scriptstyle}{\scriptscriptstyle}}
{\compscript
{\kern1.25pt\encircle{57}{21}{130}{\scriptscriptstyle{\rm Col}}\kern1.25pt}
{#1}{#2}{#3}{#4}{3.5pt}{0.5pt}{\scriptscriptstyle}{\scriptscriptstyle}}}}

\def\cirprodmsshrunk#1#2#3#4{\mathop{\mathchoice
{\ifnum\yylopstyleswitch=1\relax
\multscript
{\kern2.25pt\raise0.2pt\hbox{$\mathsurround=0pt\encircle{43}{35}{150}
{\textstyle\oldprod}$}
\vrule height11.0pt depth5.0pt width0pt\kern2.25pt}
{#1}{#2}{#3}{#4}{0pt}{1.5pt}{\displaystyle}{\scriptstyle}
\else \multscript
{\kern1.75pt\raise-0.6pt\hbox{$\mathsurround=0pt\encircle{42}{41}{120}
{\textstyle\Pi}$}
\vrule height8.0pt depth2.0pt width0pt\kern1.75pt}
{#1}{#2}{#3}{#4}{1.0pt}{0.5pt}{\textstyle}{\scriptstyle}\fi}
{\compscript
{\kern1.75pt\raise-0.6pt\hbox{$\mathsurround=0pt\encircle{42}{41}{120}
{\textstyle\Pi}$}
\vrule height8.0pt depth2.0pt width0pt\kern1.75pt}
{#1}{#2}{#3}{#4}{0.5pt}{0.5pt}{\textstyle}{\scriptstyle}}
{\compscript
{\kern1pt\raise0.6875pt\hbox{$\mathsurround=0pt\encircle{23}{20}{70}
{\scriptscriptstyle\Pi}$}
\vrule height5.47221pt depth0pt width0pt\kern1pt}
{#1}{#2}{#3}{#4}{0.5pt}{0.5pt}{\scriptstyle}{\scriptscriptstyle}}
{\compscript
{\kern1pt\encircle{23}{20}{70}{\scriptscriptstyle\Pi}\kern1pt}
{#1}{#2}{#3}{#4}{0.5pt}{0.25pt}{\scriptscriptstyle}{\scriptscriptstyle}}}}

\def\cirsummsshrunk#1#2#3#4{\mathop{\mathchoice
{\ifnum\yylopstyleswitch=1\relax
\multscript
{\kern1.75pt\raise0.2pt\hbox{$\mathsurround=0pt\encircle{47}{34}{150}
{\textstyle\oldsum}$}
\vrule height11.0pt depth5.0pt width0pt\kern1.75pt}
{#1}{#2}{#3}{#4}{0pt}{1.5pt}{\displaystyle}{\scriptstyle}
\else \multscript
{\kern2.0pt\raise-0.6pt\hbox{$\mathsurround=0pt\encircle{38}{41}{120}
{\textstyle\Sigma}$}
\vrule height8.0pt depth2.0pt width0pt\kern2.0pt}
{#1}{#2}{#3}{#4}{1.25pt}{0.5pt}{\textstyle}{\scriptstyle}\fi}
{\compscript
{\kern2.0pt\raise-0.6pt\hbox{$\mathsurround=0pt\encircle{38}{41}{120}
{\textstyle\Sigma}$}
\vrule height8.0pt depth2.0pt width0pt\kern2.0pt}
{#1}{#2}{#3}{#4}{0.75pt}{0.5pt}{\textstyle}{\scriptstyle}}
{\compscript
{\kern1.5pt\raise0.6875pt\hbox{$\mathsurround=0pt\encircle{25}{19}{70}
{\scriptscriptstyle\Sigma}$}
\vrule height5.47221pt depth0pt width0pt\kern1.5pt}
{#1}{#2}{#3}{#4}{0.5pt}{0.3pt}{\scriptstyle}{\scriptscriptstyle}}
{\compscript
{\kern1.25pt\encircle{25}{19}{70}{\scriptscriptstyle\Sigma}\kern1.25pt}
{#1}{#2}{#3}{#4}{0.5pt}{0.5pt}{\scriptscriptstyle}{\scriptscriptstyle}}}}

\def\cirLmmsshrunk#1#2#3#4{\mathop{\mathchoice
{\multscript
{\kern1.5pt\raise1.36pt\hbox{$\mathsurround=0pt\encircle{61}{27}{140}
{\scriptstyle{\rm Lm}}$}
\vrule height8.2pt depth0pt width0pt\kern1.5pt}
{#1}{#2}{#3}{#4}{3pt}{0pt}{\textstyle}{\scriptstyle}}
{\compscript
{\kern1.5pt\raise1.36pt\hbox{$\mathsurround=0pt\encircle{61}{27}{140}
{\scriptstyle{\rm Lm}}$}
\vrule height8.2pt depth0pt width0pt\kern1.5pt}
{#1}{#2}{#3}{#4}{3pt}{0pt}{\textstyle}{\scriptstyle}}
{\compscript
{\kern1.25pt\raise0.90pt\hbox{$\mathsurround=0pt\encircle{51}{20}{120}
{\scriptscriptstyle{\rm Lm}}$}
\vrule height5.47221pt depth0pt width0pt\kern1.25pt}
{#1}{#2}{#3}{#4}{2.5pt}{0.5pt}{\scriptstyle}{\scriptscriptstyle}}
{\compscript
{\kern1pt\encircle{51}{20}{120}{\scriptscriptstyle{\rm Lm}}\kern1pt}
{#1}{#2}{#3}{#4}{3pt}{0.3pt}{\scriptscriptstyle}{\scriptscriptstyle}}}}

\def\cirColmsshrunk#1#2#3#4{\mathop{\mathchoice
{\multscript
{\kern2pt\raise1.00pt\hbox{$\mathsurround=0pt\encircle{66}{26}{150}
{\scriptstyle{\rm Col}}$}
\vrule height8.33333pt depth0pt width0pt\kern2pt}
{#1}{#2}{#3}{#4}{3.5pt}{0pt}{\textstyle}{\scriptstyle}}
{\compscript
{\kern2pt\raise1.00pt\hbox{$\mathsurround=0pt\encircle{66}{26}{150}
{\scriptstyle{\rm Col}}$}
\vrule height8.33333pt depth0pt width0pt\kern2pt}
{#1}{#2}{#3}{#4}{3.5pt}{0pt}{\textstyle}{\scriptstyle}}
{\compscript
{\kern1.5pt\raise0.68pt\hbox{$\mathsurround=0pt\encircle{57}{21}{130}
{\scriptscriptstyle{\rm Col}}$}
\vrule height5.55556pt depth0pt width0pt\kern1.5pt}
{#1}{#2}{#3}{#4}{2.75pt}{0.75pt}{\scriptstyle}{\scriptscriptstyle}}
{\compscript
{\kern1.25pt\encircle{57}{21}{130}{\scriptscriptstyle{\rm Col}}\kern1.25pt}
{#1}{#2}{#3}{#4}{3.5pt}{0.5pt}{\scriptscriptstyle}{\scriptscriptstyle}}}}

\def\largecircleops{
\let\cirprodms=\cirprodmsnorm
\let\cirsumms=\cirsummsnorm
\let\cirLmms=\cirLmmsnorm
\let\cirColms=\cirColmsnorm}

\def\smallcircleops{
\let\cirprodms=\cirprodmsshrunk
\let\cirsumms=\cirsummsshrunk
\let\cirLmms=\cirLmmsshrunk
\let\cirColms=\cirColmsshrunk}

\largecircleops

%% file: mathdefs.tex
\newtheorem{theorem}{{\bf Theorem}}

\newcounter{list}		

\def\titleline#1#2#3{\hbox to \hsize{#1\hfill#2\hfill#3}}

\def\thismonth{\ifcase\month\or
 January\or February\or March\or April\or May\or June\or
 July\or August\or September\or October \or November\or December\fi{}
 \number\year}

%% file: abstract.tex
\begin{abstract}

Dataflow networks have application in various forms of stream processing, for example for parallel processing of multimedia data.  The description of dataflow graphs, including their firing behavior, is typically non-compositional and not amenable to separate compilation.  This article considers a dataflow language with a type and effect system that captures the firing behavior of actors.  This system allows definitions to abstract over actor firing rates, supporting the definition and safe composition of actor definitions where firing rates are not instantiated until a dataflow graph is launched.

\end{abstract}

%% file: intro.tex
\section{Introduction}
\label{sect:intro}

Dataflow or stream processing is becoming increasingly important, with the
growing prevalence of signal, video and audio processing, particularly on
mobile devices.  Dataflow processing is a good match with multicore and GPGPU
parallel architectures that are now prevalent on desktop computers, and will
shortly be available on consumer mobile devices.  The data parallelism of such
architectures is at least potentially a good match with the demands of stream
processing applications.  The synthesis of these architectures with stream
processing may provide a domain-specific solution to the challenge of
programming the new generations of parallel computing architectures.

Our starting point is a computational model similar to that
originally proposed by Kahn \cite{concurrency:Kah74}.  This provides for a
network of sequential \emph{actors}, each implemented in a conventional
sequential language such as C or Algol.  Actors are connected by communication
buffers on which they can send and receive data.  A key point is that actors
cannot nondeterministically select among inputs on several input channels, nor
can they test input channels for available inputs (so polling cannot be
implemented).  This restricts each actor to a completely deterministic
semantics.  The combination of implicit parallelism and deterministic
execution makes dataflow computation a good fit with some of the current
thinking of how best to successfully exploit the parallelism available in modern
multicore and GPGPU architectures, in those domains where the dataflow
paradigm is applicable.

In the embedded systems and digital signal processing community, a very useful
class of restricted Kahn networks has been identified, the so-called
\emph{synchronous dataflow (SDF)} \cite{systems:LM87:sdf} networks.  
SDF networks enable static scheduling for multi-rate applications.  
More recently, domain-specific languages such as Streamit
\cite{dataflow:Thi09:streamit} have been defined, based on the principles of
SDF, but also providing support for compiling programmer code to run on modern
parallel architectures.  

Sessional dataflow provides a framework for providing compositional
descriptions of dataflow networks \cite{dataflow:DY12:ecoop}.  A type and effect system captures
the firing behavior of actor bodies, and this information is used to
ensure that the composition of actors does not deadlock.  For simplicity, that
simple effect system did not consider variable firing rates for
actors, so for example no communication was possible within a
loop (finite loops were still useful for example for windowing
computations).  Although actors being combined could have different
rates, and adaptation of rates was part of the static checking of
actor composition, these rates were hard-coded into the software.

In this article, we consider an approach to incorporating variable
actor firing rates into dataflow descriptions.  This allows the
description of a dataflow graph to be parameterized by the firing
rates of various actors in the graph.  Actors handle variable firing rates by
performing communication in loops.  Central to this approach is the
introduction of arrays of channels and arrays of actors, and a special
form of comprehension for describing effects with this form of rate
information.  For example, in the language described in this paper,
the following loop performs downsampling on an input channel \mc{in}
by echoing every second input to the output channel \mc{out} (where
\mc{i} and \mc{o} are the type-level names for the input and output
channels, respectively):
\begin{tabbing}
\m s :~\sizeK{\inftyTy} \\
\m i :~\channelKWithDelayLimit{\channelNoDelay}{\mc{s}} \\
\m o :~\channelKWithDelayLimit{\channelNoDelay}{\mc{s}} \\
\\
\m sz :~\sizeTy{\mc{s}} \\
\m in :~\channelTy{+}{\intTy}{\mc{i}} \\
\m out :~\channelTy{-}{\intTy}{\mc{o}} \\
\m for (t,x $\in$ 1..sz) \{ \\
\Hskip\=\m \intTy\ w = \recv{\mc{in}}; \\
\>\m when (2 \divides\ x) \send{\mc{out}}{\mc{w}}; \\
\m \}
\end{tabbing}

In this example, \mc{s} is a type-level quantity that is used to model
firing rates for channels.  There are two type-level channel names,
\mc{i} and \mc{o}, whose declarations specify that there are no delays
in communication on those channels, and which have a bound of \mc{s}.  These channel names
are used in the declaration of channel variables \mc{in} and \mc{out},
respectively.  Communication on these channels is modeled at the type
level by input and output events on the corresponding type-level
channel names (\mc{i} and \mc{o}).

The size parameter \mc{s} is used in the declaration of a value-level parameter
\mc{sz}, that specifies the rate of communication on the channels.
For example, \mc{sz} is used as the bound on the loop where the
communication is performed.  
The flowstate for this loop is a sequential composition of two comprehensions
\[
\eventComp{\recvEvent{\mc{i}}}{\iterType{\mc{t}}{\numberTy{1}}{\mc{s}}}\seqAS
\eventComp{\sendEvent{\mc{o}}}{\iterType{\mc{t}}{\numberTy{1}}{\mc{s}},(\numberTy{2}\divides\mc{t})}
\]
where the type-level parameter \mc{s} models the value-level loop
bound \mc{sz}.
This can be abbreviated as a flowstate that just counts the number of
communication events on each channel:
\[  
\eventArity{\mc{s}}{\recvEvent{\mc{i}}}\seqAS
\eventArity{(\mc{s}/\numberTy{2})}{\sendEvent{\mc{o}}}.
\]

As a variation on this example, demonstrating the usefulness of
event comprehensions, we can have the code read from an array of
channels, combining several paths in a dataflow graph:
\begin{tabbing}
\m i :~\channelArrayKWithDelayLimit{\channelNoDelay}{\numberTy{1}}{\mc{s}} \\
\m in :~\channelArrayTy{+}{\intTy}{\mc{i}}{\mc{s}} \\
\m for (t,x $\in$ 1..sz) \{ \\
\Hskip\=\m \intTy\ w = \recv{\mc{in}[\mc{x}]}; \\
\>\m when (2 \divides\ x) \send{\mc{out}}{\mc{w}}; \\
\m \}
\end{tabbing}
The flowstate for this loop is a sequential composition of two comprehensions
\[
\eventComp{\recvEvent{\mc{i}[\mc{t}]}}{\iterType{\mc{t}}{\numberTy{1}}{\mc{s}}}\seqAS
\eventArity{(\mc{s}/\numberTy{2})}{\sendEvent{\mc{o}}}.
\]
In typing the code above, \mc{t} is a type-level witness for the loop
index \mc{x}.  This type witness has the kind \sizeK{\mc{s}},
reflecting that it bounded by the size parameter \mc{s}, while the
loop index has the type \indexTy{\mc{t}}.  The distinction at the type
level between
size and index variables,  \sizeTy{\mc{s}} and \indexTy{\mc{t}} respectively in
the example above, is crucial to the static analysis of buffer
sizes and actor firing rates: sizes are fixed for the execution of a
dataflow graph, while indexes obviously vary dynamically.

We consider a type system for a simple dataflow language with variable firing rates in
Sect.~\ref{sect:lang-dataflow}.  We provide an
operational semantics in Sect.~\ref{sect:sdata-semantics}.
Sect.~\ref{sect:related} considers
related work while Sect.~\ref{sect:concl} provides our conclusions.

%% file: dataflow.tex
\section{Dataflow Language}
\label{sect:types}
\label{sect:lang-dataflow}

In this section we consider a core language to prescribe
dataflow computations. We name this kernel language \Sdata.  We
describe a type system and an operational semantics for this
language.  
Our ``object language'' uses a form of session types for dataflow,
that we refer to as \emph{sessional dataflow} to express in the type
system the contracts between producers and consumers who share
message-passing channels.

\begin{figure}
\begin{figbox}
\begin{center}
\begin{tabular}{rcl}
$\Kind\in\text{Kind}$ & $::=$ & 
                              $\typeK \Mid
                                \channelK  \Mid
                                \channelArrayK{\STy}  \Mid \sizeK{\STy} $ \\
$\STy\in\text{Simple Type}$ & $::=$ & $ t \Mid
                                      \numberTy{n} 
                                      \Mid 
                                      \inftyTy
                                      \Mid
                                      (\STy_1+\STy_2) 
                                      \Mid 
                                      (\STy_1-\STy_2) 
                                      \Mid 
                                      (\STy_1*\STy_2) 
                                      \Mid 
                                      (\STy_1 / \STy_2) \Mid $ \\
                         & & 
                                      $\kw{min}(\STy_1,\STy_2)
                                      \Mid \sizeTy{\STy}
                                      \Mid \indexTy{\STy}
                                      \Mid \refTy{\STy} \Mid $ \\
  & & $(\seq{\STy}
             \funtyopr{\ActorFlowstateSym_1}{\ActorFlowstateSym_2}\STy) 
   \Mid\boolTy \Mid \intTy$ \\
$\Ty\in\text{Type}$ & $::=$ & $\STy \Mid
                              \channelTy{\polarity}{\STy}{\STy_0}  \Mid
                              \channelArrayTy{\polarity}{\STy}{\STy_0}{\STy_1}$  \\
$\channelDelayFlag\in\text{Channel Delay Flag}$ & $::=$ & 
    $\channelDelay \Mid \channelNoDelay$ \\
$\polarity\in\text{Polarity}$ & $::=$ & 
    $\posPolarity \Mid \negPolarity \Mid \pmPolarity$ \\
$\eventSym\in\text{Event}$ & $::=$ &
                $\sendEvent{t} \Mid
                  \sendEvent{t[\STy]} \Mid
                  \recvEvent{t} \Mid
                  \recvEvent{t[\STy]} $ \\
$\iterTypeSym\in\text{Type Iterator}$ & $::=$ & $
                                            (\iterType{t}{\STy_1}{\STy_2}) $
  \\
$\guardTypeSym\in\text{Type Guard}$ & $::=$ & $
                                           (\STy_1\ \relop\ \STy_2) $
  \\
$\relop\in\text{Rel Op}$ & $::=$ & \ldots \\
$\ActorFlowstateSym\in\text{Actor Flowstate}$ & $::=$ &
      $\emptyFlowstate \Mid 
                           \eventComp{\eventSym}{\seq{\iterTypeSym},\seq{\guardTypeSym}} \Mid
      (\seqFlowstate{\ActorFlowstateSym_1}{\ActorFlowstateSym_2})$ \\
$\tyenv\in\text{Type Env}$ & $::=$ &
   $\emptyEnv \Mid \tyenv,t:\Kind$ \\
$\valenv\in\text{Value Env}$ & $::=$ &
   $\emptyEnv \Mid \valenv,x:\STy \Mid \valenv,c:\Ty$ \\
$\ProcFlowstateSym\in\text{Proc Flowstate}$   & $::=$ &
    $\emptyFlowstate 
    \Mid \ActorFlowstateSym
    \Mid \fsComp{\ActorFlowstateSym}{t}{\STy_1}{\STy_2}
    \Mid (\parFlowstate{\ProcFlowstateSym_1}{\ProcFlowstateSym_2})$  \\
$\NetworkSimpleSigSym\in\text{Network sig}$ & $::=$ & 
   $\NetworkSimpleSig{\tyenv}{\valenv}{\ProcFlowstateSym}$ 
\end{tabular}
\end{center}
\end{figbox}
\caption{Abstract syntax of \protect{\Sdata} types}
\label{fig:types}
\end{figure}

The syntax of types is provided in Fig.~\ref{fig:types}.  We assume
Boolean and integer types for base types, although other types (e.g.,
floating point) could obviously be easily added.

In order to track communication rates, the type
system includes type-level names \numberTy{n} for size constants $n$, of
the form \numberTy{0}, \numberTy{1}, \numberTy{2}, \ldots.  A type
constant \numberTy{n} has the kind \sizeK{\numberTy{n}}.  A type
parameter will have a kind of the form \sizeK{\STy}, for some
type-level upper bound \STy.  In general,
the kind of a type-level numeric quantity records an upper bound on
the possible instantiations of a type parameter of that kind, and a
subkinding system allows this bound to be inflated, losing precision
in the kinds of type quantities. The special constant \inftyTy{}
represents the absence of an upper bound (the equivalent of $\top$ in
a subtyping system).

We have two types for tracking numeric quantities at the value level.
\sizeTy{\STy} represents a size parameter, fixed over the execution of
an actor.  Typically it is used to parameterize over communication rates,
or fan-in or fan-out at an actor.  \indexTy{\STy} is the type-level
representative for a loop index, which obviously does vary at
execution time.  Both types are indexed by a type-level numeric quantity of kind \sizeK{\STy}.  Our
main reason for distinguishing these two types is to prevent a loop index
being used as the bound for another loop, which would be 
useless for practical applications while complicating the analysis.  This distinction between
static and dynamic numeric quantities simplifies the
extraction of actor firing rates from behavioral types.

The language
includes type-level names for channels and channel arrays.  These are represented by type variables $t$ with
kinds of the form \channelK{} and \channelArrayK{\STy}, and are used
to index the types of values that are tracked by the type system.  So
we have a channel type \channelTy{\polarity}{t}{\STy}, where $t$ is
the type-level name for the channel (of kind \channelK), and $\STy$ the type of message
payloads that can be exchanged.  The polarity $\polarity$ allows sending or receiving on a channel.  A single
actor can only send or receive, but not both, on a channel.  In a
network, these uniplex channels in actor signatures are instantiated with shared duplex
channels that connect different actors. In the channel kind, the type
parameter $\STy_{\mathit{limit}}$ represents a type-level bound on the
number of messages that can be buffered in the channel.  
The flag $\channelDelayFlag$ indicates if messages
that should be buffered in the channel at beginning of execution of
the dataflow graph, to remove a cycle in the firing schedule by
introducing a delay.  If messages are buffered, the number of messages
to be buffered is given by the channel capacity, $\STy_{\mathit{limit}}$.

Channel array types
have the form \channelArrayTy{\polarity}{t}{\STy}{\STy_1}, where $t$ is
the type-level name for the channel array (of kind \channelArrayK{\STy_1}), $\STy$ the type of message
payloads that can be exchanged, and $\STy_1$ a bound on the size of
the array.

The language includes procedure types of the form 
$(\seq{\STy})
             \funtyopr{\ActorFlowstateSym_1}{\ActorFlowstateSym_2}\STy$.
             A procedure takes a sequence of value arguments, of type
             $\seq{\STy}$, and produces a result of type $\STy$.  In
             addition, the procedure has a latent effect, reflected by
             an actor flowstate $\ActorFlowstateSym_1$ that records the
             communications performed during the execution of this
             procedure. For mutable variables, the
             language includes references, which can be considered as
             one-element arrays.  These could be straightforwardly
             generalized to $n$-element arrays, but we use references
             for simplicity in the presentation.

An event has one of four
possible forms, two event forms for sending events and two forms
for receiving events.  For sending, the three forms are \sendEvent{c}
(sending on a channel) and \sendEvent{c[\STy]} (sending on an element of
a channel array, where $\STy$ is the type-level representative for the
index).   There are analogous event types for message
receipt: \recvEvent{c} and \recvEvent{c[\STy]}.

A flowstate for an actor $\ActorFlowstateSym$ is a 
composition of events.  In its most general form, an event in a
flowstate is
described by an \emph{event comprehension} of the form
\[
\eventComp{\eventSym}{\seq{\iterType{t}{\STy_1}{\STy_2}},\seq{\STy'\relop\STy''}}
\]
The iterators $(\iterType{t}{\STy_1}{\STy_2})$ record loops in which the event
occurred, while the guards $(\STy'\relop\STy')$ record conditions on
the occurrence of the events.  
A guard 
denotes a conditional communication, and is useful for applications such as
decimation, where an actor discards some of its input (e.g., in a
downsampler). 
We admit specific forms of this
general description of an event:
\begin{enumerate}
\item A singleton event \eventSym, that may be a communication on a
  channel or a channel array element.
\item Iterated and conditional communication over a channel \eventComp{\eventSym}{\seq{\iterTypeSym},\seq{\guardTypeSym}}, where \eventSym{} has the form
  \sendEvent{c} or \recvEvent{c} for some channel $c$.  The goal of
    the analysis is to reduce this to a single multiplicity \STy{} for
    the communication, folding guards into iterators by modifying the
    bounds, and combining the iterators.  We sometimes denote 
    \eventComp{\eventSym}{\iterType{t}{\numberTy{1}}{\STy}}, where the single
    iterator variable $t$ does not occur in \eventSym, by
    \eventArity{\STy}{\eventSym}.
We sometimes use $\eventSym$ as shorthand for
the flowstate $\eventArity{\numberTy{1}}{\eventSym}$.
\item Communication over the elements of a channel array, described by
  the comprehension \eventComp{\eventSym}{\iterType{t}{\STy_1}{\STy_2},\seq{\iterTypeSym}}, where \eventSym{} has the form
  \sendEvent{c[t]} or \recvEvent{c[t]} for some channel array $c$ of
    size $\STy$.  The iterators $\seq{\iterTypeSym}$ contribute
    additional multiplicities.  Since we can combine the additional
    iterators, if any, into a single iterator, we sometimes denote
    \eventComp{\eventSym}{\iterType{t}{\STy_1}{\STy_2},\iterType{t_0}{\STy_0}{\STy_0'}}
    by
    \eventArity{(\STy_0'-\STy_0+\numberTy{1})}{\eventComp{\eventSym}{\iterType{t}{\STy_1}{\STy_2}}},
    where $t_0$ does not occur free (as a channel array index) in the
event \eventSym.
\end{enumerate}

We must place sufficient restrictions on a guard to
ensure that it can be easily folded into a loop bound.  For this
article, we restrict guard types to be one of the following forms:
\begin{eqnarray*}
\guardTypeSym & ::= & (\STy\divides t) \Mid (t\leq
                    \STy)
\end{eqnarray*}
The first denotes a predicate asserting that the quantity $\STy$
divides the loop index $t$, while the latter asserts an upper bound on $t$.
Then we allow the following equivalences on flowstates, that fold
these conditions into iterators:
\[
\eventComp{\eventSym}{\seq{\iterTypeSym},(\iterType{t}{\STy_1}{\STy_2}),
  (\STy\divides t),\seq{\guardTypeSym}}
\equiv
\eventComp{\eventSym}{\seq{\iterTypeSym},(\iterType{t}{\numberTy{1}}{((\STy_2-\STy_1+\numberTy{1})/\STy}),\seq{\guardTypeSym}}
\]
\[
\eventComp{\eventSym}{\seq{\iterTypeSym},(\iterType{t}{\STy_1}{\STy_2}),(t\leq \STy),\seq{\guardTypeSym}}
\equiv
\eventComp{\eventSym}{\seq{\iterTypeSym},(\iterType{t}{\STy_1}{\kw{min}(\STy,\STy_2)}),\seq{\guardTypeSym}}
\]

We only allow conditions on communication in the case where
communication is on a channel rather than a channel array, and in this
case the actual range of values of the iteration variable is not
important, since we are only counting number of occurrences of the
communication event in a firing.

A dataflow network has  a \emph{network
  signature} $\NetworkSimpleSigSym$, which has three
parts:
\begin{enumerate}
\item A type environment \tyenv{} that binds type-level
  representatives for channels, channel arrays and sizes.
\item A value environment \valenv{} that captures information about the
  shared communication channels, using bindings of the
  form $(c:\Ty)$, as well as size parameters for the network description. 
\item The \emph{flowstate} of a network $\ProcFlowstateSym$ records its expected firing
  behavior.  This is described by the parallel composition of the
  flowstates of the actors in the network.
\end{enumerate}

\begin{figure}
\begin{figbox}
\begin{center}
\begin{tabular}{rcll}
$\val\in\text{Values}$ & $::=$ & 
   \multicolumn{2}{l}{$\kw{true} \Mid \kw{false} \Mid n \Mid c \Mid x \Mid \sizeExp{\val}$ } \\
   & $\mid$ & \multicolumn{2}{l}{$\indexExp{\val} \Mid \lambda \seq{x}:\seq{\STy}{.}
     \lambdar{\ActorFlowstateSym_1}{\ActorFlowstateSym_2}
      \Exp$ } \\
$\Exp\in\text{Expr}$ & $::=$ & 
  $\lambda \seq{x}:\seq{\STy}{.}
     \lambdar{\ActorFlowstateSym_1}{\ActorFlowstateSym_2}
      \Exp$  & Abstraction \\
  & $\mid$ &
  $\Exp(\seq{\Exp})$  & Application \\
  & $\mid$ &
   $\kw{let}\ x=\Exp_1\ \kw{in}\ \Exp_2$ & Bind variable \\
   & $\mid$ &
   $\kw{if}\ \Exp\ \kw{then}\ \Exp_1\ \kw{else}\ \Exp_2$  
        & Conditional  \\
   & $\mid$ &
   $\kw{when}\ \Exp_1\ \kw{do}\ \Exp_2$  
        & Cond Comm  \\
   & $\mid$ &
   $\kw{for}\ (t,x\in n..\Exp)\ \Exp_2$  & Finite Loop \\
    & $\mid$ & 
    $\newRef{\Exp}$  & New reference \\
    & $\mid$ & 
    $\derefExp{\Exp}$  & Dereference \\
    & $\mid$ & 
    $\assignExp{\Exp_1}{\Exp_2}$  & Assignment \\
    & $\mid$ & 
    $\sizeExp{\Exp}$  & Size constant \\
    & $\mid$ & 
    $\fromSizeExp{\Exp}$  & Size projection \\
    & $\mid$ & 
    $\indexExp{\Exp}$  & Loop index \\
    & $\mid$ & 
    $\fromIndexExp{\Exp}$  & Index projection \\
    & $\mid$ & 
    $\recv{\chanLab}$, $\recv{\chanLab[\Exp_0]}$ & Receive a message \\
    & $\mid$ & 
    $\send{\chanLab}{\Exp}$, $\send{\chanLab[\Exp_0]}{\Exp}$ & Send a message \\
$\Proc\in\text{Proc}$ & $::=$ & \multicolumn{2}{l}{$\stopProc \Mid {\Exp} \Mid
\actorComp{\Exp}{t}{x}{\val_1}{\val_2}
\Mid (\Proc_1\parsym\Proc_2)$ } \\
$\NetworkSimpleSym\in\text{Network}$ & $::=$ &
    $\NetworkSimple{\tyenv}{\valenv}{\ProcFlowstateSym}{\Proc}$
\end{tabular}
\end{center}
\end{figbox}
\caption{Abstract syntax of \protect{\Sdata} expressions and processes}
\label{fig:expressions}
\end{figure}

Fig.~\ref{fig:expressions} provides the abstract syntax for programs in
\Sdata.  Values include Booleans (\kw{true} and \kw{false}) and
integers $n$.   Atomic
values also include names $c$ (for channels and breakpoints), and variables  $x$.  Our
language is a basic expressional language, with functions and
call-by-value evaluation.  An abstraction, of the form 
$\lambda \seq{x}{:}\seq{\STy}{.}
      \lambdar{\ActorFlowstateSym_1}{\ActorFlowstateSym_2}
      \Exp$, abstracts over simple value parameters.
      For now we disallow value-level abstraction over type parameters,
      such as those for singleton types for channels,
      as well as numeric quantities, in order to avoid aliasing
      issues.  We allow for abstraction over such parameters in the
      network graph as a whole, and the instantiation of the network
      ensures that no aliases are introduced.  
      The latent flowstate $\ActorFlowstateSym_1$  for the procedure body
      $\Exp$ (i.e., the communications
      it offers) is provided
      as an annotation.  As with the function type, this records the
      communication performed in the function $\ActorFlowstateSym_1$.
An application 
$\Exp(\seq{\Exp})$ 
denotes the application of
a procedure to 
value level
parameters $\seq{\Exp}$.  
A \kw{let}
construct, which can be read as a combination of abstraction and
application, 
 binds a variable in a local context. 

A conditional allows dispatching on a Boolean value.  The type rules
require that both branches in the conditional have identical
flowstates.  To record
conditional communication in an actor, the \kw{when} construct relates
Boolean conditions to communication events in the flowstate.

The finite loop construct binds two local parameters: $x$, the index
variable for the loop, and $t$, a type-level parameter for the loop
index.  The latter, in combination with channel array references in
communication events, is used to record communication behavior in a
loop.
A parameter of size type is created by the constructor
$\sizeExp{n}\in\sizeTy{\numberTy{n}}$, and is deconstructed using the
accessor $\fromSizeExp{\Exp}$, providing access to the underlying
integer bound.  Similar operations are available for constructing loop
index values $\indexExp{n}\in\indexTy{\numberTy{n}}$ and projecting
the loop index out of this value, $\fromIndexExp{\Exp}$.

There are two operations for receiving messages, receiving on a
channel or on a channel array element, and similarly two operations
for sending messages.
 It is instructive
that the channel reference is always a name and never a variable.  In this
account, we are not yet considering a facility for transmitting the ability to
send or receive on a channel, as is found in varying degrees in the
pi-calculus.  The reason is again to avoid issues with channel aliasing, which
would subvert flowstate checking on usage of channels.

The type system is formulated using judgements of the following forms:
\[
        \begin{array}{ll}
        \tyenvj{\tyenv}  & \text{Type environment}  \\
        \valenvj{\tyenv}{\valenv}  & \text{Value environment} \\
        \kindj{\tyenv}{\Kind}  & \text{Kind} \\
        \tyj{\tyenv}{\STy}{\Kind}  & \text{Type} \\
        \eventj{\tyenv}{\eventSym}  & \text{Event} \\
        \flowj{\tyenv}{\ActorFlowstateSym}  & \text{Flowstate} \\
        \expj{\tyenv}{\valenv}{\Exp}{\STy}{\ActorFlowstateSym}  & \text{Expression} \\
        \procj{\tyenv}{\valenv}{\Proc}{\ProcFlowstateSym}  & \text{Process} \\
        \end{array}
\]

\begin{figure}[t]
\begin{figbox}
\begin{center}
\begin{minipage}{0.3\textwidth}
\infer[\inflab{TyEnv Empty}]{
  \tyenvj{\emptyEnv}
}{
  \mbox{ }
}
\end{minipage}
\Hskip
\begin{minipage}{0.3\textwidth}
\infer[\inflab{TyEnv Extend}]{
  \tyenvj{\tyenv,t:\Kind}
}{
  \tyenvj{\tyenv}
  \Hskip
  \kindj{\tyenv}{\Kind}
}
\end{minipage}

\begin{minipage}{0.3\textwidth}
\infer[\inflab{Kind Size}]{
  \kindj{\tyenv}{\sizeK{\STy}}
}{
  \tyj{\tyenv}{\STy}{\sizeK{\_}}
}
\end{minipage}

\begin{minipage}{0.3\textwidth}
\infer[\inflab{Kind Chan}]{
  \kindj{\tyenv}{\channelK}
}{
  \tyenvj{\tyenv}
  &
  \tyj{\tyenv}{\STy_{\mathit{limit}}}{\sizeK{\_}}
  &
  \tyj{\tyenv}{\STy_{\mathit{init}}}{\sizeK{\STy_{\mathit{limit}}}}
}
\end{minipage}
\Hskip
\begin{minipage}{0.3\textwidth}
\infer[\inflab{Kind Chan Array}]{
  \kindj{\tyenv}{\channelArrayK{\STy}}
}{
  \tyj{\tyenv}{\STy}{\sizeK{\_}}
  &
  \tyj{\tyenv}{\STy_{\mathit{limit}}}{\sizeK{\_}}
  &
  \tyj{\tyenv}{\STy_{\mathit{init}}}{\sizeK{\STy_{\mathit{limit}}}}
}
\end{minipage}

\begin{minipage}{0.3\textwidth}
\infer[\inflab{Size Refl}]{
  \sizeleqj{\tyenv}{\STy}{\STy}
}{
  \tyj{\tyenv}{\STy}{\sizeK{\_}}
}
\end{minipage}
\Hskip
\begin{minipage}{0.3\textwidth}
\infer[\inflab{Size Trans}]{
  \sizeleqj{\tyenv}{\STy_1}{\STy_3}
}{
  \sizeleqj{\tyenv}{\STy_1}{\STy_2}
  &
  \sizeleqj{\tyenv}{\STy_2}{\STy_3}
}
\end{minipage}

\begin{minipage}{0.3\textwidth}
\infer[\inflab{Size Infty}]{
  \sizeleqj{\tyenv}{\STy}{\inftyTy}
}{
  \tyj{\tyenv}{\STy}{\sizeK{\_}}
}
\end{minipage}
\Hskip
\begin{minipage}{0.3\textwidth}
\infer[\inflab{Size Bound}]{
  \sizeleqj{\tyenv}{\STy}{\STy_0}
}{
  \tyj{\tyenv}{\STy}{\sizeK{\STy_0}}
}
\end{minipage}
\Hskip
\begin{minipage}{0.3\textwidth}
\infer[\inflab{Size Num}]{
  \sizeleqj{\tyenv}{\numberTy{m}}{\numberTy{n}}
}{
  \tyenvj{\tyenv}
  &
  m\leq n
}
\end{minipage}

\begin{minipage}{0.3\textwidth}
\infer[\inflab{Ty Size}]{
  \tyj{\tyenv}{\numberTy{n}}{\sizeK{\numberTy{n}}}
}{
  \tyenvj{\tyenv}
}
\end{minipage}
\Hskip
\begin{minipage}{0.3\textwidth}
\infer[\inflab{Ty Var}]{
  \tyj{\tyenv}{t}{\Kind}
}{
  \tyenvj{\tyenv}
  &
  (t:\Kind)\in\tyenv
}
\end{minipage}
\Hskip
\begin{minipage}{0.3\textwidth}
\infer[\inflab{Ty Infty}]{
  \tyj{\tyenv}{\inftyTy}{\sizeK{\inftyTy}}
}{
  \tyenvj{\tyenv}
}
\end{minipage}
\Hskip
\begin{minipage}{0.3\textwidth}
\infer[\inflab{Ty Chan}]{
  \tyj{\tyenv}{\channelTy{\polarity}{\STy_1}{\STy_2}}{\typeK}
}{
  \tyj{\tyenv}{\STy_1}{\channelK}
  & 
  \tyj{\tyenv}{\STy_2}{\typeK}
}
\end{minipage}
\Hskip
\begin{minipage}{0.3\textwidth}
\infer[\inflab{Ty Chan Array}]{
  \tyj{\tyenv}{\channelArrayTy{\polarity}{\STy_1}{\STy_2}{\STy_3}}{\typeK}
}{
  \begin{array}{c}
  \tyj{\tyenv}{\STy_1}{\channelArrayK{\STy_3}}
  \Hskip
  \tyj{\tyenv}{\STy_2}{\typeK}
    \end{array}
}
\end{minipage}
\end{center}
\end{figbox}
\caption{Type Environments, Kinds and Types}
\label{fig:env-types}
\end{figure}

\begin{figure}[t]
\begin{figbox}
\begin{center}
\begin{minipage}{0.3\textwidth}
\infer[\inflab{FS Send}]{
  \eventj{\tyenv}{\sendEvent{\STy}}
}{
  \tyj{\tyenv}{\STy}{\channelK}
}
\end{minipage}
\Hskip
\begin{minipage}{0.3\textwidth}
\infer[\inflab{FS Recv}]{
  \eventj{\tyenv}{\recvEvent{\STy}}
}{
  \tyj{\tyenv}{\STy}{\channelK}
}
\end{minipage}

\begin{minipage}{0.3\textwidth}
\infer[\inflab{FS Array Send}]{
  \eventj{\tyenv}{\sendEvent{\STy[\STy_1]}}
}{
  \tyj{\tyenv}{\STy}{\channelArrayK{\STy_0}}
  &
  \sizeleqj{\tyenv}{\STy_1}{\STy_0}
}
\end{minipage}
\Hskip
\begin{minipage}{0.3\textwidth}
\infer[\inflab{FS Array Recv}]{
  \eventj{\tyenv}{\recvEvent{\STy[\STy_1]}}
}{
  \tyj{\tyenv}{\STy}{\channelArrayK{\STy_0}}
  &
  \sizeleqj{\tyenv}{\STy_1}{\STy_0}
}
\end{minipage}

\begin{minipage}{0.3\textwidth}
\infer[\inflab{FS Iter}]{
  \iterj{\tyenv}{\iterType{t}{\STy_1}{\STy_2}}
}{
  \tyj{\tyenv}{\STy_1}{\sizeK{\_}}
  &
  \tyj{\tyenv}{\STy_1}{\sizeK{\_}}
}
\end{minipage}
\Hskip
\begin{minipage}{0.3\textwidth}
\infer[\inflab{FS Gd Div}]{
  \guardj{\tyenv}{(\STy_1\divides\STy_2)}
}{
  \tyj{\tyenv}{\STy_1}{\sizeK{\_}}
  &
  \tyj{\tyenv}{\STy_1}{\sizeK{\_}}
}
\end{minipage}
\Hskip
\begin{minipage}{0.3\textwidth}
\infer[\inflab{FS Gd Bnd}]{
  \guardj{\tyenv}{(\STy_1\leq\STy_2)}
}{
  \tyj{\tyenv}{\STy_1}{\sizeK{\_}}
  &
  \tyj{\tyenv}{\STy_1}{\sizeK{\_}}
}
\end{minipage}

\begin{minipage}{0.3\textwidth}
\infer[\inflab{FS Comp}]{
  \flowj{\tyenv}{\eventComp{\eventSym}{\seq{\iterTypeSym},\seq{\guardTypeSym}}}
}{
  \seq{\iterTypeSym}=\seq{\iterType{t}{\STy_1}{\STy_2}}
  &
  \eventj{\tyenv,\seq{t}:\seq{\sizeK{\STy_2}}}{\eventSym}
  &
  \seq{\iterj{\tyenv}{\iterTypeSym}}
  &
  \seq{\guardj{\tyenv}{\guardTypeSym}}
}
\end{minipage}

\begin{minipage}{0.3\textwidth}
\infer[\inflab{FS Empty}]{
  \flowj{\tyenv}{\emptyFlowstate}
}{
  \tyenvj{\tyenv}
}
\end{minipage}
\Hskip
\begin{minipage}{0.3\textwidth}
\infer[\inflab{FS Seq}]{
  \flowj{\tyenv}{\seqFlowstate{\ActorFlowstateSym_1}{\ActorFlowstateSym_2}}
}{
  \flowj{\tyenv}{\ActorFlowstateSym_1}
  &
  \flowj{\tyenv}{\ActorFlowstateSym_2}
}
\end{minipage}
\end{center}
\end{figbox}
\caption{Flowstates}
\label{fig:flowstates}
\end{figure}

\begin{figure}[t]
\begin{figbox}
\begin{center}
\begin{minipage}{0.3\textwidth}
\infer[\inflab{ValEnv Empty}]{
  \valenvj{\tyenv}{\emptyEnv}
}{
  \tyenvj{\tyenv}
}
\end{minipage}
\Hskip
\begin{minipage}{0.3\textwidth}
\infer[\inflab{ValEnv Extend Var}]{
  \valenvj{\tyenv}{\valenv,x:\STy}
}{
  \valenvj{\tyenv}{\valenv}
  \Hskip
  \tyj{\tyenv}{\STy}{\typeK}
}
\end{minipage}

\begin{minipage}{0.3\textwidth}
\infer[\inflab{ValEnv Extend Name}]{
  \valenvj{\tyenv}{\valenv,c:\Ty}
}{
  \valenvj{\tyenv}{\valenv}
  \Hskip
  \tyj{\tyenv}{\Ty}{\typeK}
}
\end{minipage}
\Hskip
\begin{minipage}{0.3\textwidth}
\infer[\inflab{Val Int}]{
  \expj{\tyenv}{\valenv}{n}{\intTy}{\actorpcr{\emptyFlowstate}{\ActorFlowstateSym}}
}{
  \valenvj{\tyenv}{\valenv}
  \Hskip
  \flowj{\tyenv}{\ActorFlowstateSym}
}
\end{minipage}

\begin{minipage}{0.3\textwidth}
\infer[\inflab{Val True}]{
  \expj{\tyenv}{\valenv}{\kw{true}}{\boolTy}{\actorpcr{\emptyFlowstate}{\ActorFlowstateSym}}
}{
  \valenvj{\tyenv}{\valenv}
  \Hskip
  \flowj{\tyenv}{\ActorFlowstateSym}
}
\end{minipage}
\Hskip
\begin{minipage}{0.3\textwidth}
\infer[\inflab{Val False}]{
  \expj{\tyenv}{\valenv}{\kw{false}}{\boolTy}{\actorpcr{\emptyFlowstate}{\ActorFlowstateSym}}
}{
  \valenvj{\tyenv}{\valenv}
  \Hskip
  \flowj{\tyenv}{\ActorFlowstateSym}
}
\end{minipage}

\begin{minipage}{0.3\textwidth}
\infer[\inflab{Val Var}]{
  \expj{\tyenv}{\valenv}{x}{\STy}{\actorpcr{\emptyFlowstate}{\ActorFlowstateSym}}
}{
  \valenvj{\tyenv}{\valenv}
  &
  (x:\STy)\in\valenv
  \Hskip
  \flowj{\tyenv}{\ActorFlowstateSym}
}
\end{minipage}
\end{center}
\end{figbox}
\caption{Value Environments and Values}
\label{fig:env-values}
\end{figure}

The type rules for environments, kinds and types are provided in
Fig.~\ref{fig:env-types}.  A ``size kind'' \sizeK{\STy} is indexed by
an upper bound on numeric types of that kind.  This in turn gives rise
to a subkinding relationship \sizeleqj{\tyenv}{\STy_1}{\STy_1} (that
can be read as a synonym for
\tyj{\tyenv}{\STy_1}{\sizeK{\STy_2}}), formalized in
Fig.~\ref{fig:env-types}.  Note that whereas we allow subsumption on
size kinds, we do not allow subtyping on size types (of the form
\sizeTy{\STy} or \indexTy{\STy}, for some witness of size kind
\sizeK{\_}), because size estimates in types, and particularly in flowstates, are required to be precise.

Fig.~\ref{fig:flowstates} provides formation rules for flowstates,
while Fig.~\ref{fig:env-values} provides type rules for value
environments and values.

\begin{figure*}
\begin{figbox*}
\begin{center}
\[
\infer[\inflab{Val Let}]{
  \expj{\tyenv}{\valenv}
        {(\kw{let}\ x=\Exp_1\ \kw{in}\ \Exp_2)}
        {\STy_2}
        {\actorpcr{(\seqFlowstate{\ActorFlowstateSym_1}{\ActorFlowstateSym_2})}
          {\ActorFlowstateSym}}
  }{
   \begin{array}{c}
  \expj{\tyenv}{\valenv}
        {\Exp_1}
        {\STy_1}
        {\actorpcr{\ActorFlowstateSym_1}{(\seqFlowstate{\ActorFlowstateSym_2}\ActorFlowstateSym{})}}
   \Hskip
  \expj{\tyenv}{(\valenv,x:\STy_1)}
        {\Exp_2}
        {\STy_2}
        {\actorpcr{\ActorFlowstateSym_2}{\ActorFlowstateSym}}
   \end{array}
  }
\]
\[
\infer[\inflab{Val Abs}]{
  \expj{\tyenv}{\valenv}
        {(\lambda \seq{x}:\seq{\STy_1}{.} \lambdar{\ActorFlowstateSym_1}{\ActorFlowstateSym_2}
                                                          \Exp)}
        {(\seq{\STy_1}\funtyopr{\ActorFlowstateSym_1}{\ActorFlowstateSym_2}\STy_2)}
        {\actorpcr{\emptyFlowstate}{\ActorFlowstateSym}}
  }{
   \begin{array}{c}
  \expj{\tyenv}{\valenv,\seq{x}:\seq{\STy_1}}
        {\Exp}
        {\STy_2}
        {\actorpcr{\ActorFlowstateSym_1}{\ActorFlowstateSym_2}}
   \end{array}
  }
\]
\[
\infer[\inflab{Val App}]{
  \expj{\tyenv}{\valenv}
        {\Exp(\Exp_1,\ldots,\Exp_k)}
        {\STy}
        {\actorpcr{(\ActorFlowstateSym_0\seqAS \ActorFlowstateSym_1\seqAS \ldots\seqAS \ActorFlowstateSym_k\seqAS \ActorFlowstateSym)}
                      {\ActorFlowstateSym'}}
  }{
   \begin{array}{c}
  \expj{\tyenv}{\valenv}
        {\Exp_0}
        {((\STy_1,\ldots,\STy_k)
          \funtyopr{\ActorFlowstateSym}{{\ActorFlowstateSym'}}\STy)}
        {\actorpcr{\ActorFlowstateSym_0}{(\ActorFlowstateSym_1\seqAS \ldots\seqAS \ActorFlowstateSym_k\seqAS \ActorFlowstateSym\seqAS \ActorFlowstateSym')}}
     \\
  \expj{\tyenv}{\valenv}
        {\Exp_i}
        {\STy_i}
        {\actorpcr{\ActorFlowstateSym_i}{(\ActorFlowstateSym_{i+1}\seqAS \ldots\seqAS \ActorFlowstateSym_k\seqAS \ActorFlowstateSym\seqAS \ActorFlowstateSym')}}
     \ \text{for}\ 
                    i=1,\ldots,k
  \end{array}
  }
\]
\[
\begin{minipage}{0.3\textwidth}
\infer[\inflab{Val Cond}]{
  \expj{\tyenv}{\valenv}
        {(\kw{if}\ \Exp_0\ \kw{then}\ \Exp_1\ \kw{else}\ \Exp_2)}
        {\STy}
        {\actorpcr{(\seqFlowstate{\ActorFlowstateSym_0}{\ActorFlowstateSym})}
          {\ActorFlowstateSym'}}
  }{
    \begin{array}{c}
  \expj{\tyenv}{\valenv}
        {\Exp_0}
        {\boolTy}
        {\actorpcr{\ActorFlowstateSym_0}{(\seqFlowstate{\ActorFlowstateSym}{\ActorFlowstateSym'})}}
   \\
  \expj{\tyenv}{\valenv}
        {\Exp_1}
        {\STy}
        {\actorpcr{\ActorFlowstateSym}{\ActorFlowstateSym'}}
   \Hskip
  \expj{\tyenv}{\valenv}
        {\Exp_2}
        {\STy}
        {\actorpcr{\ActorFlowstateSym}{\ActorFlowstateSym'}}
     \end{array}
  }
\end{minipage}
\]
\[
\begin{minipage}{0.3\textwidth}
\infer[\inflab{Val Ref}]{
  \expj{\tyenv}{\valenv}
        {\newRef{\Exp}}
        {\refTy{\STy}}
        {\actorpcr{\ActorFlowstateSym_0}{\ActorFlowstateSym}}
  }{
  \expj{\tyenv}{\valenv}
        {\Exp}
        {\STy}
        {\actorpcr{\ActorFlowstateSym_0}{\ActorFlowstateSym}}
 }
\end{minipage}
\]
\[
\begin{minipage}{0.3\textwidth}
\infer[\inflab{Val Deref}]{
  \expj{\tyenv}{\valenv}
        {\derefExp{\Exp}}
        {\STy}
        {\actorpcr{\ActorFlowstateSym_0}{\ActorFlowstateSym}}
  }{
  \expj{\tyenv}{\valenv}
        {\Exp}
        {\refTy{\STy}}
        {\actorpcr{\ActorFlowstateSym_0}{\ActorFlowstateSym}}
 }
\end{minipage}
\]
\[
\begin{minipage}{0.3\textwidth}
\infer[\inflab{Val Assign}]{
  \expj{\tyenv}{\valenv}
        {\assignExp{\Exp_1}{\Exp_2}}
        {\STy}
        {\actorpcr{(\ActorFlowstateSym_1\seqAS \ActorFlowstateSym_2)}{\ActorFlowstateSym}}
  }{
  \expj{\tyenv}{\valenv}
        {\Exp_1}
        {\refTy{\STy}}
       {\actorpcr{(\ActorFlowstateSym_1}{(\ActorFlowstateSym_2\seqAS \ActorFlowstateSym)}}
   \Hskip
  \expj{\tyenv}{\valenv}
        {\Exp_2}
        {\STy}
       {\actorpcr{\ActorFlowstateSym_2}{\ActorFlowstateSym}}
  }
\end{minipage}
\]
\[
\infer[\inflab{Val Eq}]{
  \expj{\tyenv}{\valenv}
        {\Exp}
        {\STy}
        {\actorpcr{\ActorFlowstateSym_1}{\ActorFlowstateSym_2}}
}{
  \begin{array}{c}
  \expj{\tyenv}{\valenv}
        {\Exp}
        {\STy}
        {\actorpcr{\ActorFlowstateSym_1'}{\ActorFlowstateSym_2'}}
   \Hskip
   \flowj{\tyenv}{\ActorFlowstateSym_1}
   \Hskip
   \flowj{\tyenv}{\ActorFlowstateSym_2}
   \\
    \ActorFlowstateSym_1\equiv\ActorFlowstateSym_1'
    \Hskip
    \ActorFlowstateSym_2\equiv\ActorFlowstateSym_2'
    \end{array}
}
\]
\end{center}
\end{figbox*}
\caption{\protect{\Sdata}: Core Expressions}
\label{fig:sdata-type-rules-core}
\end{figure*}

\begin{figure*}
\begin{figbox*}
\begin{center}
\[
\begin{minipage}{0.3\textwidth}
\infer[\inflab{Val For}]{
  \expj{\tyenv}{\valenv}
        {(\kw{for}\ (t,x\in 1..\Exp_1)\ \Exp_2)}
        {\STy}
        {(\actorpcr{(\ActorFlowstateSym_0\seqAS \eventComp{\ActorFlowstateSym_1}{\iterType{t}{\numberTy{1}}{\STy}})}{\ActorFlowstateSym_2})}
  }{
    \begin{array}{c}
  \expj{\tyenv}{\valenv}
        {\Exp_1}
        {\sizeTy{\STy_1}}
        {\actorpcr{\ActorFlowstateSym_0}{(\eventComp{\ActorFlowstateSym_1}{\iterType{t}{\numberTy{1}}{\STy}}\seqAS \ActorFlowstateSym_2)}}
   \\
  \expj{\tyenv,t:\sizeK{\STy_1}}{\valenv,x:\indexTy{t}}
        {\Exp_2}
        {\STy_2}
        {(\actorpcr{\ActorFlowstateSym_1}{(\eventComp{\subst {\ActorFlowstateSym_1}{t_0}{t}}{\iterType{t_0}{(t+\numberTy{1})}{\STy}})\seqAS \ActorFlowstateSym_2})}
     \end{array}
  }
\end{minipage}
\]

\[
\begin{minipage}{0.3\textwidth}
\infer[\inflab{Val When}]{
  \expj{\tyenv}{\valenv}
        {(\kw{when}\ \Exp_1\relop\Exp_2\ \kw{do}\ \Exp_3)}
        {\STy}
        {(\actorpcr{(\ActorFlowstateSym_1\seqAS \ActorFlowstateSym_2\seqAS \eventComp{\ActorFlowstateSym_3}{\STy_1\relop\STy_2})}{\ActorFlowstateSym})}
  }{
    \begin{array}{c}
  \expj{\tyenv}{\valenv}
        {\Exp_1}
        {\STy_1}
        {\actorpcr{\ActorFlowstateSym_1}{(\ActorFlowstateSym_2\seqAS \eventComp{\ActorFlowstateSym_3}{\STy_1\relop\STy_2}\seqAS \ActorFlowstateSym)}}
   \\
  \expj{\tyenv}{\valenv}
        {\Exp_2}
        {\STy_2}
        {\actorpcr{\ActorFlowstateSym_2}{(\eventComp{\ActorFlowstateSym_3}{\STy_1\relop\STy_2}\seqAS \ActorFlowstateSym)}}
   \\
   \wfguard{\STy_1}{\relop}{\STy_2}
      \Hskip
  \expj{\tyenv}{\valenv}
        {\Exp}
        {\STy}
        {(\actorpcr{\ActorFlowstateSym_3}{\ActorFlowstateSym})}
     \end{array}
  }
\end{minipage}
\]

\[
\begin{minipage}{0.3\textwidth}
\infer[\inflab{Val Size}]{
  \expj{\tyenv}{\valenv}
        {\sizeExp{n}}
        {\sizeTy{\numberTy{n}}}
        {\actorpcr{\emptyFlowstate}{\ActorFlowstateSym}}
  }{
  \valenvj{\tyenv}{\valenv}
  \Hskip
  \flowj{\tyenv}{\ActorFlowstateSym}
 }
\end{minipage}
\]
\[
\begin{minipage}{0.3\textwidth}
\infer[\inflab{Val Int}]{
  \expj{\tyenv}{\valenv}
        {\fromSizeExp{\Exp}}
        {\intTy}
        {\actorpcr{\ActorFlowstateSym_1}{\ActorFlowstateSym_2}}
  }{
  \expj{\tyenv}{\valenv}
        {\Exp}
        {\sizeTy{\STy}}
        {\actorpcr{\ActorFlowstateSym_1}{\ActorFlowstateSym_2}}
 }
\end{minipage}
\]

\[
\begin{minipage}{0.3\textwidth}
\infer[\inflab{Val Index}]{
  \expj{\tyenv}{\valenv}
        {\indexExp{n}}
        {\indexTy{\numberTy{n}}}
        {\actorpcr{\emptyFlowstate}{\ActorFlowstateSym}}
  }{
  \valenvj{\tyenv}{\valenv}
  \Hskip
  \flowj{\tyenv}{\ActorFlowstateSym}
 }
\end{minipage}
\]
\[
\begin{minipage}{0.3\textwidth}
\infer[\inflab{Val FromIndex}]{
  \expj{\tyenv}{\valenv}
        {\fromIndexExp{\Exp}}
        {\intTy}
        {\actorpcr{\ActorFlowstateSym_1}{\ActorFlowstateSym_2}}
  }{
  \expj{\tyenv}{\valenv}
        {\Exp}
        {\indexTy{\STy}}
        {\actorpcr{\ActorFlowstateSym_1}{\ActorFlowstateSym_2}}
 }
\end{minipage}
\]

\[
\begin{minipage}{0.3\textwidth}
\infer[\inflab{Val Send}]{
  \expj{\tyenv}{\valenv}
        {\send{\chanLab}{\Exp}}
        {\STy}
        {\actorpcr{(\seqFlowstate{\ActorFlowstateSym_1}{\sendEvent{\STy_0}})}{\ActorFlowstateSym_2}}
  }{
  \expj{\tyenv}{\valenv}
        {\Exp}
        {\STy}
        {\actorpcr{\ActorFlowstateSym_1}{(\sendEvent{\STy_0}\seqAS \ActorFlowstateSym_2)}}
     &
    (\chanLab:\channelTy{\polarity}{\STy_0}{\STy})\in\valenv
    &
    \polarity\in\{\negPolarity,\pmPolarity\}
 }
\end{minipage}
\]
\[
\begin{minipage}{0.3\textwidth}
\infer[\inflab{Val Receive}]{
  \expj{\tyenv}{\valenv}
        {\recv{\chanLab}}
        {\STy}
        {\actorpcr{\recvEvent{\STy_0}}{\ActorFlowstateSym}}
  }{
    \valenvj{\tyenv}{\valenv}
   \Hskip
    (\chanLab:\channelTy{\polarity}{\STy_0}{\STy})\in\valenv
    \Hskip
    \flowj{\tyenv}{\ActorFlowstateSym}
    \Hskip
    \polarity\in\{\posPolarity,\pmPolarity\}
  }
\end{minipage}
\]
\[
\begin{minipage}{0.3\textwidth}
\infer[\inflab{Val Send Array}]{
  \expj{\tyenv}{\valenv}
        {\send{\chanLab[\Exp_0]}{\Exp}}
        {\STy}
        {\actorpcr{(\ActorFlowstateSym_0 \seqAS \ActorFlowstateSym_1\seqAS \sendEvent{\STy_1[\STy_0]})}{\ActorFlowstateSym_2}}
  }{
   \begin{array}{c}
    (\chanLab:\channelArrayTy{\polarity}{\STy_1}{\STy}{\STy_0'})\in\valenv
    \Hskip
    \polarity\in\{\negPolarity,\pmPolarity\}
     \\
  \expj{\tyenv}{\valenv}
        {\Exp_0}
        {\indexTy{\STy_0}}
        {\actorpcr{\ActorFlowstateSym_0}{(\ActorFlowstateSym_1\seqAS \sendEvent{\STy_1[\STy_0]}\seqAS \ActorFlowstateSym_2)}}
     \\
  \expj{\tyenv}{\valenv}
        {\Exp_1}
        {\STy_1}
        {\actorpcr{\ActorFlowstateSym_1}{(\sendEvent{\STy_1[\STy_0]}\seqAS \ActorFlowstateSym_2)}}
        \Hskip
        \sizeleqj{\tyenv}{\STy_0}{\STy_0'}
    \end{array}
 }
\end{minipage}
\]
\[
\begin{minipage}{0.3\textwidth}
\infer[\inflab{Val Recv Array}]{
  \expj{\tyenv}{\valenv}
        {\recv{\chanLab[\Exp_0]}}
        {\STy}
        {\actorpcr{(\ActorFlowstateSym_0\seqAS \recvEvent{\STy_0[\STy_1]})}{\ActorFlowstateSym}}
  }{
    \begin{array}{c}
    (\chanLab:\channelArrayTy{\polarity}{\STy_1}{\STy}{\STy_0'})\in\valenv
    \Hskip
    \polarity\in\{\posPolarity,\pmPolarity\}
     \\
  \expj{\tyenv}{\valenv}
        {\Exp_0}
        {\indexTy{\STy_0}}
        {\actorpcr{\ActorFlowstateSym_0}{(\recvEvent{\STy_1[\STy_0]}\seqAS \ActorFlowstateSym_1)}}
      \Hskip
        \sizeleqj{\tyenv}{\STy_0}{\STy_0'}
    \end{array}
  }
\end{minipage}
\]
\end{center}
\end{figbox*}
\caption{\protect{\Sdata}: Dataflow Expressions}
\label{fig:sdata-type-rules-exp}
\end{figure*}

\begin{figure*}
\begin{figbox*}
\begin{center}
\[
  \infer[\inflab{Proc Empty}]{
    \procj{\tyenv}{\valenv}{\stopProc}{\emptyFlowstate}
    }{
    }
\]
\[
  \infer[\inflab{Proc Exp}]{
    \procj{\tyenv}{\valenv}{\Exp}{\ActorFlowstateSym}
    }{
    \expj{\tyenv}{\valenv}{\Exp}{\Ty}{\ActorFlowstateSym}
    }
\]
\[
  \infer[\inflab{Proc Comp}]{
    \procj{\tyenv}{\valenv}{\actorComp{\Exp}{t}{x}{n}{\val}}{\fsComp{\ActorFlowstateSym}{t}{\numberTy{n}}{\Ty}}
    }{
    \valj{\tyenv}{\valenv}{\val}{\sizeTy{\Ty}}
    &
    \expj{\tyenv,t:\sizeK{\Ty}}{\valenv,x:\indexTy{t}}{\Exp}{\Ty}{\ActorFlowstateSym}
    }
\]
\[
  \infer[\inflab{Proc Par}]{
    \procj{\tyenv}{\valenv}{(\Proc_1\parsym\Proc_2)}{\parFlowstate{\ProcFlowstateSym_1}{\ProcFlowstateSym_2}}
    }{
    \procj{\tyenv}{\valenv}{\Proc_1}{\ProcFlowstateSym_1}
    &
    \procj{\tyenv}{\valenv}{\Proc_2}{\ProcFlowstateSym_2}
    }
\]
\end{center}
\end{figbox*}
\caption{\protect{\Sdata}: Processes}
\label{fig:sdata-proc-type-rules}
\end{figure*}

Fig.~\ref{fig:sdata-type-rules-core} and
Fig.~\ref{fig:sdata-type-rules-exp} provide type rules for
value-level expressions in the language.
The main type rules for dataflow computation are provided in
Fig.~\ref{fig:sdata-type-rules-exp}.  
The \kw{for} construct is key, since it allows communication within a loop, as
represented by the flowstate $\ActorFlowstateSym_1$ for each
iteration.  The loop bound has a type-level witness type $\STy_1$, so the
flowstate for the entire loop is $(\eventComp{\ActorFlowstateSym_1}{\iterType{t}{\numberTy{1}}{\STy}})$.
Within the loop, the index variable has type $\indexTy{t}$, where $t$
is an index type parameter of kind $\sizeK{\STy_1}$, bounded above
by the loop bound.  $\ActorFlowstateSym_2$ is the flowstate for the
remaining computation after the loop.  Within the loop, thie remaining
flowstate after each iteration represented by the flowstate expression
$(\eventComp{\subst
  {\ActorFlowstateSym_1}{t_0}{t}}{\iterType{t_0}{(t+\numberTy{1})}{\STy}})$,
where $t_0$ is a new variable introduced to count the remaining
iterations.

The expression
\eventComp{\ActorFlowstateSym}{\iterType{t}{\STy_1}{\STy_2}} is a
metafunction, generalizing the event comprehension introduced earlier
from events to flowstates, defined by:
\begin{eqnarray*}
\eventComp{\emptyFlowstate}{\iterType{t}{\STy_1}{\STy_2}} & = &
                                                                \emptyFlowstate \\
\eventComp{\ActorFlowstateSym_1\seqAS \ActorFlowstateSym_2}{\iterType{t}{\STy_1}{\STy_2}} & = &
\eventComp{\ActorFlowstateSym_1}{\iterType{t}{\STy_1}{\STy_2}}\seqAS 
\eventComp{\ActorFlowstateSym_2}{\iterType{t}{\STy_1}{\STy_2}} \\
\eventComp{\eventComp{\eventSym}{\seq{\iterTypeSym},\seq{\guardTypeSym}}}{\iterType{t}{\STy_1}{\STy_2}} & = &
                                  \eventComp{\eventSym}{(\seq{\iterTypeSym},\iterType{t}{\STy_1}{\STy_2}),\seq{\guardTypeSym}} 
 \end{eqnarray*}
There is an important assumption in the second case of this
definition, where we distribute a flowstate comprehension over the
joining of two flowstates.  The assumption is that we are not tracking
causality within an actor, so we are free to reorder communications
within that actor.  For example a loop that inputs on one channel
$c_1$ and
outputs on another $c_2$ would have the flowstate
\eventComp{\recvEvent{c_1}\seqAS \sendEvent{c_2}}{\iterType{t}{\numberTy{1}}{\numberTy{n}}},
which normalizes to
$(\eventComp{\recvEvent{c_1}}{\iterType{t}{\numberTy{1}}{\numberTy{n}}}\seqAS \eventComp{\sendEvent{c_2}}{\iterType{t}{\numberTy{1}}{\numberTy{n}}})$,
which we abbreviate as 
$(\eventArity{\numberTy{n}}{\recvEvent{c_1}}\seqAS \eventArity{\numberTy{n}}{\sendEvent{c_2}})$.
The normalized form loses the causality between a receive and send on a single
loop iteration.  We rely on a global check of the composition of the
actors to detect any causal cycles in the firing of a dataflow graph,
relying on the fact that we do not have abstraction over the structure
of the dataflow graph (beyond the eliding of internal causality within
an actor).  A global check stratifies the actors based on
communication dependencies, and ensures there are no cycles where an
actor's inputs depend on its own outputs (unless there is a delay in
the channel).

The constructs for sizes, that are used to track capacity bounds and
communication rates, allow an integer literal to be wrapped as a value $\sizeExp{n}$
whose type \sizeTy{\numberTy{n}} reflects the integer quantity.  The
destructor $\fromSizeExp{\Exp}$ allows this size parameter to be projected
to an integer, with $\fromSizeExp{\sizeExp{n}}$ evaluating to $n$.
Similar constructs are available for dynamic numeric quantities (loop indices).

There are four communication primitives: two for sending and two for
receiving.  Each sending and receiving operation has a variant for
communicating on a single channel or communicating on an element of a
channel array.  Each of these primitives gives rise to one of the four
communication events tracked by the flowstate: $\sendEvent{c}$,
$\recvEvent{c}$, $\sendEvent{c[\STy]}$ and $\recvEvent{c[\STy]}$.  For
example, a loop that outputs one value on each element of a channel
array would have the form:
\[
\kw{for}\ (t,x\in 1..\Exp_1)\ \send{c[x]}{\Exp_2}
\]
Assume $\Exp_1$ has type $\sizeTy{\STy}$, where $\STy$ is some
type-level size of kind $\sizeK{\_}$, e.g., $\numberTy{n}$ of kind
$\sizeK{\numberTy{n}}$.  The loop variable has type $\indexTy{t}$,
where the loop type variable $t$ has kind $\sizeK{\STy}$, reflecting
the type-level bound on the number of iterations.
Each iteration of the loop has a flowstate $\sendEvent{c[t]}$, and the
entire loop has the flowstate
$\eventComp{\sendEvent{c[t]}}{\iterType{t}{\numberTy{1}}{\STy}}$.

Conditional communication is performed using the \kw{when} construct.
This tests a condition and adds a guard to the flowstate for the body
of the \kw{when}.  The condition is restricted so that it can only be
used to refine static bounds, and not introduce a dependency in
computation bounds on dynamic quantities (such as a loop index).  The
well-formedness condition \wfguard{\STy_1}{\relop}{\STy_2}
on the types of the two values being
compared enforces this restriction, where $\STy_1$ and $\STy_2$ are
the types of the expressions being compared, and $\relop$ is the
relational operator:
\begin{eqnarray*}
\wfguard{\STy_1}{\divides}{\STy_2} & \myiff &
                                              \STy_1=\sizeTy{\_} \
                                              \mathrm{and}\
                                              \STy_1=\indexTy{\_} \\
\wfguard{\STy_1}{\leq}{\STy_2} & \myiff &
                                              \STy_1=\indexTy{\_} \
                                              \mathrm{and}\
                                              \STy_1=\sizeTy{\_} 
\end{eqnarray*}

As with iterators, we define a metafunction that distributes guards
over flowstates.
The expression
\eventComp{\ActorFlowstateSym}{\STy_1\relop\STy_2} is a
metafunction, generalizing the event comprehension introduced earlier
from events to flowstates, defined by:
\begin{eqnarray*}
\eventComp{\emptyFlowstate}{\STy_1\relop\STy_2} & = &
                                                                \emptyFlowstate \\
\eventComp{\ActorFlowstateSym_1\seqAS \ActorFlowstateSym_2}{\STy_1\relop\STy_2} & = &
\eventComp{\ActorFlowstateSym_1}{\STy_1\relop\STy_2}\seqAS 
\eventComp{\ActorFlowstateSym_2}{\STy_1\relop\STy_2} \\
\eventComp{\eventComp{\eventSym}{\seq{\iterTypeSym},\seq{\guardTypeSym}}}{\STy_1\relop\STy_2} & = &
                                  \eventComp{\eventSym}{\seq{\iterTypeSym},(\seq{\guardTypeSym},\STy_1\relop\STy_2)}
 \end{eqnarray*}

A \emph{dataflow network}
$\NetworkSym=\NetworkSimple{\tyenv}{\valenv}{\ProcFlowstateSym}{\Proc}$
is a composition of sequential actors
under certain conditions.  The composition $\Proc$
contains two form of actor bindings:
\begin{enumerate}
\item A simple actor binding of the form \Exp.
\item An \emph{actor comprehension}, an actor array binding of the form
  \actorComp{\Exp}{t}{x}{\val_1}{\val_2}, that represents an array of actor, all of the same
  definition $\Exp$, and with each actor provided with its index via the
  parameter $x$ when it is initialized.  This index ranges over the
  interval $\{\val_1,\ldots,\val_2\}$.  An actor comprehension is the
  parallel equivalent of a loop for processing an array of channels.
\end{enumerate}

In order to ensure the well-formedness of a network, we define some
global restrictions based on the flowstates for the actors.  We
formulate these restrictions using the following additional judgement
forms:
\begin{enumerate}
\item
  \livefsj{\tyenv}{\ProcFlowstateSym_1}{\ProcFlowstateSym_2}{\ProcFlowstateSym_1'}{\ProcFlowstateSym_2'}:
  Determines if it is possible to evolve from an initial flowstate
  $\ProcFlowstateSym_1$ to a flowstate $\ProcFlowstateSym_1'$, where
  communications represented by $\ProcFlowstateSym_2$ have already
  occurred, and leaving an updated record of communications
  $\ProcFlowstateSym_2'$.  The ``records of communications'' correspond
  to send events that are preconditions for receive events (data
  cannot be read until it written), and in the case of channels with
  delays, receive events that are preconditions for send events (data
  on a channel with a delay cannot be written until the data written
  in the previous firing cycle has been read). 
\item \detfsj{\tyenv}{\ProcFlowstateSym}: Determines that there is a
  single sending actor and single receiving actor for each channel
  (and it is not the same actor sending and receiving on a channel).
The rules for this judgement form are provided in FIg.~\ref{fig:det-fs}.
\end{enumerate}

\begin{figure}
\begin{figbox}
\[
\infer[\inflab{FS Prog Empty}]{
    \livefsj{\tyenv}{\emptyFlowstate}{\ProcFlowstateSym}{\emptyFlowstate}{\ProcFlowstateSym}
}{
}
\]
\[
\infer[\inflab{FS Prog Par}]{
    \livefsj{\tyenv}{(\parFlowstate{\ProcFlowstateSym_1}{\ProcFlowstateSym_2})}{\ProcFlowstateSym_3}{(\parFlowstate{\ProcFlowstateSym_1'}{\ProcFlowstateSym_2})}{\ProcFlowstateSym_3'}
}{
    \livefsj{\tyenv}{\ProcFlowstateSym_1}{\ProcFlowstateSym_3}{\ProcFlowstateSym_1'}{\ProcFlowstateSym_3'}
}
\]
\[
\infer[\inflab{FS Prog Prod}]{
    \livefsj{\tyenv}{(\seqFlowstate{\eventComp{\eventSym}{\seq{\iterTypeSym},\seq{\guardTypeSym}}}{\ActorFlowstateSym})}{\ProcFlowstateSym}{\ActorFlowstateSym}{(\parFlowstate{\eventComp{\eventSym}{\seq{\iterTypeSym},\seq{\guardTypeSym}}}{\ProcFlowstateSym})}
}{
    \mathit{Producer}(\tyenv,\eventSym)
}
\]
\[
\infer[\inflab{FS Prog Cons}]{
    \livefsj{\tyenv}{(\seqFlowstate{\eventComp{\eventSym}{\seq{\iterTypeSym},\seq{\guardTypeSym}}}{\ActorFlowstateSym})}{(\parFlowstate{\eventComp{\eventCompl{\eventSym}}{\seq{\iterTypeSym},\seq{\guardTypeSym}}}{\ProcFlowstateSym})}{\ActorFlowstateSym}{\ProcFlowstateSym}
}{
    \mathit{Consumer}(\tyenv,\eventSym)
}
\]
\[
\infer[\inflab{FS Prog Cong}]{
    \livefsj{\tyenv}{\ProcFlowstateSym_1}{\ProcFlowstateSym_2}{\ProcFlowstateSym_3}{\ProcFlowstateSym_4}
}{
          \livefsj{\tyenv}{\ProcFlowstateSym_1'}{\ProcFlowstateSym_2'}{\ProcFlowstateSym_3'}{\ProcFlowstateSym_4'}
          &
          \ProcFlowstateSym_i\equiv\ProcFlowstateSym_i'\ \mathrm{for}\ i=1,\ldots,4
}
\]
\end{figbox}
\caption{Progress Conditions for Flowstate}
\label{fig:live-fs}
\end{figure}

 The rules for the first
  judgement form are provided in Fig.~\ref{fig:live-fs}.  These divide
  events into two broad categories: ``producer'' events and
  ``consumer'' events, defined by these predicates:
\begin{eqnarray*}
\mathit{Producer}(\tyenv,\sendEvent{t}) & \myiff & 
                                                 \tyenv(t)=\channelKWithDelay{\channelNoDelay} \\
\mathit{Producer}(\tyenv,\sendEvent{t[\STy]}) & \myiff & 
                                                 \tyenv(t)=\channelKWithDelay{\channelNoDelay} \\
\mathit{Producer}(\tyenv,\recvEvent{t}) & \myiff & 
                                                 \tyenv(t)=\channelKWithDelay{\channelDelay} \\
\mathit{Producer}(\tyenv,\recvEvent{t[\STy]}) & \myiff & 
                                                 \tyenv(t)=\channelKWithDelay{\channelDelay} \\
\mathit{Consumer}(\tyenv,\sendEvent{t}) & \myiff & 
                                                 \tyenv(t)=\channelKWithDelay{\channelDelay} \\
\mathit{Consumer}(\tyenv,\sendEvent{t[\STy]}) & \myiff & 
                                                 \tyenv(t)=\channelKWithDelay{\channelDelay} \\
\mathit{Consumer}(\tyenv,\recvEvent{t}) & \myiff & 
                                                 \tyenv(t)=\channelKWithDelay{\channelNoDelay} \\
\mathit{Consumer}(\tyenv,\recvEvent{t[\STy]}) & \myiff & 
                                                 \tyenv(t)=\channelKWithDelay{\channelNoDelay} 
\end{eqnarray*}

In other words, $\mathit{Producer}(\tyenv,\eventSym)$ is true if
$\eventSym$ corresponds to a communication event that is a
precondition for another communication event in this firing cycle
(sending on a channel with no delay, or receiving on a channel with a
delay).  $\mathit{Consumer}(\tyenv,\eventSym)$ is true if
$\eventSym$ corresponds to a communication event that relied on a
preceding communication event in this firing cycle.

In the rules in Fig.~\ref{fig:live-fs}, the \inflab{FS Prog Prod} rule
corresponds to a flowstate event that produces the communication to
enable a subsequent event, e.g., the sending of a message that will
later be consumed, on a channel with no delay:
\[
    \livefsj{\tyenv}{(\seqFlowstate{\eventComp{\eventSym}{\seq{\iterTypeSym},\seq{\guardTypeSym}}}{\ActorFlowstateSym})}{\ProcFlowstateSym}{\ActorFlowstateSym}{(\parFlowstate{\eventComp{\eventSym}{\seq{\iterTypeSym},\seq{\guardTypeSym}}}{\ProcFlowstateSym})}.
\]
  The \inflab{FS Prog
  Cons} rule corresponds to a flowstate event that consumes the result
of this communication later in the computation:
\[
    \livefsj{\tyenv}{(\seqFlowstate{\eventComp{\eventSym}{\seq{\iterTypeSym},\seq{\guardTypeSym}}}{\ActorFlowstateSym})}{(\parFlowstate{\eventComp{\eventCompl{\eventSym}}{\seq{\iterTypeSym},\seq{\guardTypeSym}}}{\ProcFlowstateSym})}{\ActorFlowstateSym}{\ProcFlowstateSym}.
\]
  This latter rule
makes use of the notion of the \emph{complement of an event}, defined
by:
\begin{eqnarray*}
\eventCompl{\sendEvent{t}} & = & \recvEvent{t} \\
\eventCompl{\sendEvent{t[\STy]}} & = & \recvEvent{t[\STy]} \\
\eventCompl{\recvEvent{t}} & = & \sendEvent{t} \\
\eventCompl{\recvEvent{t[\STy]}} & = & \sendEvent{t[\STy]} 
\end{eqnarray*}

\begin{figure}
\begin{figbox}
\[
\infer[\inflab{FS Det Empty}]{
    \detfsj{\tyenv}{\emptyFlowstate}
}{
}
\]
\[
\infer[\inflab{FS Det Actor}]{
    \detfsj{\tyenv}{\ActorFlowstateSym}
}{
}
\]
\[
\infer[\inflab{FS Det Par}]{
    \detfsj{\tyenv}{(\parFlowstate{\ProcFlowstateSym_1}{\ProcFlowstateSym_2})}
}{
  \begin{array}{c}
    \detfsj{\tyenv}{\ProcFlowstateSym_1}
    \Hskip
    \detfsj{\tyenv}{\ProcFlowstateSym_2}
    \\
    \mathit{inchans}(\ProcFlowstateSym_1)\cap\mathit{inchans}(\ProcFlowstateSym_2)=\emptyset 
    \Hskip
    \mathit{outchans}(\ProcFlowstateSym_1)\cap\mathit{outchans}(\ProcFlowstateSym_2)=\emptyset 
   \end{array}
}
\]
\end{figbox}
\caption{Determinism Conditions for Flowstate}
\label{fig:det-fs}
\end{figure}

The rules for the  \detfsj{\tyenv}{\ProcFlowstateSym} judgement form
are provided in Fig.~\ref{fig:det-fs}.  These rules make use of
metafunctions that extract the input and output channels that
processes communicate on, as reflected in the flowstate:
\begin{eqnarray*}
\mathit{inchans}(\sendEvent{t},\seq{\iterTypeSym},\seq{\guardTypeSym}) & = & \emptyset \\
\mathit{inchans}(\sendEvent{t[\STy]},\seq{\iterTypeSym},\seq{\guardTypeSym}) & = & \emptyset \\
\mathit{inchans}(\recvEvent{t},\seq{\iterTypeSym},\seq{\guardTypeSym}) & = & \{t\} \\
\mathit{inchans}(\recvEvent{t[\STy]},\seq{\iterTypeSym},\seq{\guardTypeSym}) & = & \left\{ 
                                            \begin{array}{ll}
                                              \{t[\numberTy{k}]\} & \text{if}\
                                                      \STy=\numberTy{k},
                                              \ \text{some}\ n\\
                                              \{t[\numberTy{m}],\ldots,t[\numberTy{n}]\} & \text{if}\
                                                      \exists t_0 . \STy=t_0\
                                                            \text{and}\
                                                            \seq{\iterTypeSym}=(t_0\leftarrow\numberTy{m}..\numberTy{n})\ 
                                                            \text{and}\
                                                            \seq{\guardTypeSym}=\varepsilon\\
                                              \{t[\numberTy{m}..t_1]\} & \text{if}\
                                                      \exists t_0 . \STy=t_0\
                                                            \text{and}\
                                                            \seq{\iterTypeSym}=(t_0\leftarrow\numberTy{m}..t_1)\ 
                                                            \text{and}\
                                                            \seq{\guardTypeSym}=\varepsilon\\
                                            \end{array}  \right. \\
\mathit{inchans}(\emptyFlowstate) & = & \emptyset \\
\mathit{inchans}(\eventComp{\eventSym}{\seq{\iterTypeSym},\seq{\guardTypeSym}})
                                  & = & \mathit{inchans}(\eventSym ,\seq{\iterTypeSym},\seq{\guardTypeSym}) \\
\mathit{inchans}(\seqFlowstate{\ActorFlowstateSym_1}{\ActorFlowstateSym_2})
                                  & = & 
                                        \mathit{inchans}(\ActorFlowstateSym_1) 
                                        \cup\mathit{inchans}(\ActorFlowstateSym_2) \\
\mathit{inchans}(\parFlowstate{\ProcFlowstateSym_1}{\ProcFlowstateSym_2})
                                  & = & 
                                        \mathit{inchans}(\ProcFlowstateSym_1) 
                                        \cup\mathit{inchans}(\ProcFlowstateSym_2) \\
\mathit{outchans}(\sendEvent{t},\seq{\iterTypeSym},\seq{\guardTypeSym}) & = & \{t\} \\
\mathit{outchans}(\sendEvent{t[\STy]},\seq{\iterTypeSym},\seq{\guardTypeSym}) & = & \left\{ 
                                           \begin{array}{ll}
                                              \{t[\numberTy{k}]\} & \text{if}\
                                                      \STy=\numberTy{k},
                                              \ \text{some}\ n\\
                                              \{t[\numberTy{m}],\ldots,t[\numberTy{n}]\} & \text{if}\
                                                      \exists t_0 . \STy=t_0\
                                                            \text{and}\
                                                            \seq{\iterTypeSym}=(t_0\leftarrow\numberTy{m}..\numberTy{n})\ 
                                                            \text{and}\
                                                            \seq{\guardTypeSym}=\varepsilon\\
                                              \{t[\numberTy{m}..t_1]\} & \text{if}\
                                                      \exists t_0 . \STy=t_0\
                                                            \text{and}\
                                                            \seq{\iterTypeSym}=(t_0\leftarrow\numberTy{m}..t_1)\ 
                                                            \text{and}\
                                                            \seq{\guardTypeSym}=\varepsilon\\
                                            \end{array}  \right. \\
\mathit{outchans}(\recvEvent{t},\seq{\iterTypeSym},\seq{\guardTypeSym}) & = & \emptyset \\
\mathit{outchans}(\recvEvent{t[\STy]},\seq{\iterTypeSym},\seq{\guardTypeSym}) & = & \emptyset \\
\mathit{outchans}(\emptyFlowstate) & = & \emptyset \\
\mathit{outchans}(\eventComp{\eventSym}{\seq{\iterTypeSym},\seq{\guardTypeSym}})
                                  & = & \mathit{outchans}(\eventSym ,\seq{\iterTypeSym},\seq{\guardTypeSym}) \\
\mathit{outchans}(\seqFlowstate{\ActorFlowstateSym_1}{\ActorFlowstateSym_2})
                                  & = & 
                                        \mathit{outchans}(\ActorFlowstateSym_1) 
                                        \cup\mathit{outchans}(\ActorFlowstateSym_2) \\
\mathit{outchans}(\parFlowstate{\ProcFlowstateSym_1}{\ProcFlowstateSym_2})
                                  & = & 
                                        \mathit{outchans}(\ProcFlowstateSym_1) 
                                        \cup\mathit{outchans}(\ProcFlowstateSym_2) 
\end{eqnarray*}

The $\mathit{inchans}$ metafunction computes the set of channels on
which a part of a dataflow network performs inputs (as reflected by
the flowstate inferred for that network). 
The obvious complication is
that for channel arrays. 
 In general, we assume an
event comprehension of the form 
\[
  \eventComp{\recvEvent{t[t_0]}}{(\iterType{t_0}{\numberTy{m}}{\STy}),\seq{\guardTypeSym}}
\]
for some filtering conditions $\guardTypeSym$.  We assume that these
filtering conditions are empty for communication on an array of
channels: There is communication on every channel in the array.  We would expect that
the lower bound $\numberTy{m}$ is $\numberTy{1}$, while the upper
bound $\STy$ may be a type parameter.  However we also use the type
system to type check intermediate configurations in the next section,
and in this case the lower bound may be greater than $1$.  In that
case the upper bound will be instantiated to a number $\numberTy{n}$,
and the equivalence rules allow the comprehension to be unrolled to a
collection of singletons $\recvEvent{t[\numberTy{k}]}$ for
$k=m,\ldots,n$.  We use heap typing to compute a flowstate that
reflects communications that have been performed in this firing
cycle.  Then for an event comprehension of the form
$\eventComp{\recvEvent{t[t_0]}}{(\iterType{t_0}{\numberTy{m}}{\numberTy{n}})}$,
  where $m>1$, the preceding inputs on channel $t_0$ will be reflected
  by singleton events $\recvEvent{t[\numberTy{k}]}$ for
  $k=1,\ldots,m-1$.  

Similar remarks apply for the metafunction that computes the range of
channels on which a subnet performs outputs.

We distinguish two cases when computing the channels on which a subnet
performs input or output:
\begin{enumerate}
\item For the case where an agent communicates on an
  element of an array $t$, at index \numberTy{k}, we represent this by the
  array element $t[\numberTy{k}]$.  This corresponds to the case where
  the upper bound on an event comprehension is instantiated to a
  number $\numberTy{n}$, and the equivalence rules unroll the event
  comprehension to a collection of communications on the elements of
  the channel array.
\item For the case where an agent communicates on a range of elements
  of an array $t$, as given by the iterator 
  \iterType{t_0}{\numberTy{m}}{\STy}, we have two cases:
  \begin{enumerate}
    \item If the upper bound $\STy$ is instantiated
      (e.g. $\numberTy{n}$), then we compute the channels that are
      communicated over to be the same as those resulting from the
      unfolding of the comprehension, $\{t[\numberTy{m}],\ldots,t[\numberTy{n}]\}$
\item  The other case is where the upper bound on an event
  comprehension is not yet instantiated.  Let this supper bound be
  $t_1$, then we compute the range of channels as
  $t[\numberTy{m}..t_1]$ (where $m=1$ in this case, since no loop
  unrolling happens before instantiation of the upper bound).
\end{enumerate}
\end{enumerate}

\section{Operational Semantics}
\label{sect:sdata-semantics}

\begin{figure}
\begin{figbox}
\begin{eqnarray*}
\Heap \in\text{Heap} & ::= & \emptyHeap \Mid
   \heapBind{l}{\val} \Mid
   \heapBind{c}{\Buffer} \Mid
   \heapBind{c}{\arrayVal{\seq{\Buffer}}} \Mid
   \heapJoin{\Heap_1}{\Heap_2}  \\
\HeapTy \in\text{Heap Type} & ::= & \emptyHeap \Mid
   \heapTyBind{l}{\STy} \Mid 
   \heapTyBind{c}{\Ty} \Mid 
   \heapTyJoin{\HeapTy_1}{\HeapTy_2} \\
\Buffer\in\text{Buffer} & ::= & \emptyBuff{k} \Mid \buffCell{k}{\val} \Mid 
     \joinBuff{k}{\Buffer_1}{\Buffer_2} \\
\ConfigSym \in\text{Config} & ::= & 
      \Config{\Proc}{\Heap}
\end{eqnarray*}
\end{figbox}
\caption{Configurations in \protect{\Sdata}}
\label{fig:configurations-Sdata}
\end{figure}
 
We provide a heap-based semantics that
binds channels to message buffers on the heap.
Message buffers $\Buffer$ hold the values transmitted between
  actors on shared channels.  A message buffer is simply a sequence, ensuring
  FIFO delivery, where \joinBuff{k}{\_}{\_} is the operation for appending
  buffers.  We assume that buffers have bounded size, provided by a
  parameter $k$ in the constructors and in the buffer type; the
  constructor operations are undefined for the case where the
  resulting buffer is larger than the maximum size.  We denote the
  number of items in a buffer by \sizeBuff{\Buffer}, and the maximum
  size of a buffer by \maxsizeBuff{\Buffer}.
We write $[\val_1,\val_2,\ldots,\val_m]_k$ as an abbreviation for
  $\joinBuff{k}{\buffCell{k}{\val_1}}{\joinBuff{k}{\buffCell{k}{\val_2}}{\joinBuff{k}{\ldots}{\buffCell{k}{\val_m}}}}$,
  where $m\leq k$.
  We use $\consBuff{k}{\val}{\Buffer}$ to denote
  $\joinBuff{k}{\buffCell{k}{\val}}{\Buffer}$.  We use $\BufferTy{k}{\STy}$ to denote the
type of a buffer that contains values of type \STy.  These buffer types are not
first class, since buffers are handled by the compiler.  

There are three types of values stored on the heap: simple values (for
reference cells), buffers (for communication channels), and arrays of
buffers (for arrays of communication channels.
An array value is a tuple of
the form $\Array=\arrayVal{\val_0,\ldots,\val_{k-1}}$.  We denote 
$\maxsizeBuff{\Array}=k$.  We denote array lookup by
$\Array(i)=\val_i$, for $0\leq i<k$, and array update by
$\assignArrayVal{\Array}{i}{\val})(j)=\arrayVal{\val_0,\ldots,\val,\ldots,\val_{k-1}}$,
replacing the $i$th element of the array. 

In order to reason about correctness, we define typing relations for
heaps, using the judgement forms:
\[  \begin{array}{ll}
        \heapj{\tyenv}{\valenv}{\Heap}{\ProcFlowstateSym} &
        \text{Heap} \\
        \buffj{\tyenv}{\valenv}{\Buffer}{\ProcFlowstateSym} &
        \text{Buffer} 
        \end{array}  
\]
The type system not only ensures that values stored in a buffer have
the correct type, but also that the buffer has sufficient items to
satisfy communications between actors sharing that buffer.

For evaluating expressions, mutable base type variables are bound to locations
$l$, and these must be dereferenced.  This dereferencing is performed by the
operation of applying the heap to a value, $\Heap(\val)$, defined by:
\begin{eqnarray*}
\Heap(l) & = & \val\ \mathrm{if}\ \heapBind{l}{\val}\in\Heap \\
\Heap(c) & = & \Buffer\ \mathrm{if}\ \heapBind{c}{\Buffer}\in\Heap \\
\Heap(c[\indexExp{i}]) & = & \Buffer_i\ \mathrm{if}\ \heapBind{c}{\arrayVal{\seq{\Buffer}}}\in\Heap
\end{eqnarray*}

The semantics is defined using a collection of reduction relations:
\begin{center}
\begin{tabular}{rl}
Reduction of expressions: & 
  $\Config{\Exp_1}{\Heap_1} \iredExp
   \Config{\Exp_2}{\Heap_2}$
  and \\
& 
  $\Config{\Exp_1}{\Heap_1} \redExp{\valenv}{\eventSym}
   \Config{\Exp_2}{\Heap_2}$ \\
Reduction of processes: & $\Config{\Proc_1}{\Heap_1} \iredProc{a}
   \Config{\Proc_2}{\Heap_2}$
   and \\
& 
    $\Config{\Proc_1}{\Heap_1} \redProc{\valenv}{\eventSym}
   \Config{\Proc_2}{\Heap_2}$ \\
Reduction of flowstates:  &
                       $\airedTyj{\tyenv}{\CausalitySetSym}{\ProcFlowstateSym_1}{\ProcFlowstateSym_2}$
                       and \\
  & $\aredTyj{\tyenv}{\CausalitySetSym}{\ProcFlowstateSym_1}{\eventSym}{\ProcFlowstateSym_2}$
\end{tabular}
\end{center}

A reduction of expressions of the form $\Config{\Exp_1}{\Heap_1} \iredExp
   \Config{\Exp_2}{\Heap_2}$ denotes an internal reduction, while a reduction
   of the form 
   $\Config{\Exp_1}{\Heap_1} \redExp{\valenv}{\eventSym}
   \Config{\Exp_2}{\Heap_2}$ denotes a reduction that involves a communication
   event 
  $\eventSym$.
We write 
$\Config{\Exp_1}{\Heap_1} \optRedExp{\valenv}{\eventSym}
   \Config{\Exp_2}{\Heap_2}$ to generically denote a reduction that may be
   either internal or involve a communication event.  
   Similar remarks hold for
    reduction  of processes.

The reduction relation for flowstates is perhaps surprising, and reflects the use of flowstate:  Types
themselves evolve during computation, since they are abstract process
descriptions for the underlying sequential program. 
The reduction relation for flowstates is defined in
Fig.~\ref{fig:type-reduction-rules}.

\begin{figure}
\begin{figbox}
\begin{center}
\[
\infer[\text{\sc Red For True}]{
  \Config{(\kw{for}\ (t,x\in m..\sizeExp{n})\ \Exp)}{\Heap}
  \iredExp
  \Config{(\subst{\Exp}{\indexExp{m}}{x};\Exp')}{\Heap}
  }{
    m\leq n
    \Hskip
    \Exp'=(\kw{for}\ (t,x\in (m+1)..n)\ \Exp)
  }
\]
\[
\infer[\text{\sc Red For False}]{
  \Config{(\kw{for}\ (t,x\in m..\sizeExp{n})\ \Exp)}{\Heap}
  \iredExp
  \Config{0}{\Heap}
  }{
    m>n
  }
\]
\[
\infer[\text{\sc Red When True}]{
  \Config{(\kw{when}\ \val_1\relop\val_2\ \kw{do}\ \Exp)}{\Heap}
  \iredExp
  \Config{\Exp}{\Heap}
  }{
    \val_1\relop\val_2
  }
\]
\[
\infer[\text{\sc Red When False}]{
  \Config{(\kw{when}\ \val_1\relop\val_2\ \kw{do}\ \Exp)}{\Heap}
  \iredExp
  \Config{0}{\Heap}
  }{
    \neg(\val_1\relop\val_2)
  }
\]
\[
\infer[\text{\sc Red From Size}]{
  \Config{\fromSizeExp{\sizeExp{m}}}{\Heap}
  \iredExp
  \Config{m}{\Heap}
  }{
  }
\]
\[
\infer[\text{\sc Red From Index}]{
  \Config{\fromIndexExp{\indexExp{m}}}{\Heap}
  \iredExp
  \Config{m}{\Heap}
  }{
  }
\]
\[\infer[\text{\sc Red Send}]{
  \Config{\send{\negChan{c}}{\val}}{\Heap}
  \redExp{\valenv}{\sendEvent{\negChan{t}}}
  \Config{0}{\Heap'}
  }{
    \begin{array}{c}
   k=\maxsizeBuff{\Heap(c)}
      \Hskip
    \sizeBuff{\Heap(c)} < k
    \Hskip
    \Heap'=\Heap[\heapBind{c}{\joinBuff{k}{\Heap(c)}{\buffCell{k}{\val}}}]
      \\
      (c:\channelTy{\polarity}{t}{\STy})\in\valenv
    \end{array}
  }
\]
\[
\infer[\text{\sc Red Receive}]{
  \Config{\recv{\posChan{c}}}{\Heap}
  \redExp{\valenv}{\recvEvent{\posChan{t}}}
  \Config{\val}{\Heap'}
  }{
  \begin{array}{c}
    \Heap(c)=\joinBuff{k}{\buffCell{k}{\val}}{\Buffer}
    \Hskip
    \Heap'=\Heap[\heapBind{c}{\Buffer}]
      \Hskip
      (c:\channelTy{\polarity}{t}{\STy})\in\valenv
  \end{array}
  }
\]
\[\infer[\text{\sc Red Send Array}]{
  \Config{\send{\negChan{c}[\val_0]}{\val}}{\Heap}
  \redExp{\valenv}{\sendEvent{\negChan{t}[\numberTy{m}]}}
  \Config{0}{\Heap'}
  }{
    \begin{array}{c}
   k=\maxsizeBuff{\Heap(c[\val_0])}
      \Hskip
    \sizeBuff{\Heap(c[\val_0])} < k
    \Hskip
      \val_0=\indexExp{m}
      \\
    \Heap'=\Heap[\heapBind{c[\val_0]}{\joinBuff{k}{\Heap(c[\val_0])}{\buffCell{k}{\val}}}]
      \Hskip
      (c:\channelArrayTy{\polarity}{t}{\STy}{\STy_0})\in\valenv
    \end{array}
  }
\]
\[
\infer[\text{\sc Red Receive Array}]{
  \Config{\recv{\posChan{c}[\val_0]}}{\Heap}
  \redExp{\valenv}{\recvEvent{\posChan{t}[\numberTy{m}]}}
  \Config{\val}{\Heap'}
  }{
  \begin{array}{c}
    \Heap(c[\val_0])=\joinBuff{k}{\buffCell{k}{\val}}{\Buffer}
    \Hskip
    \Heap'=\Heap[\heapBind{c[\val_0]}{\Buffer}]
    \Hskip
      \val_0=\indexExp{m}
      \\
      (c:\channelArrayTy{\polarity}{t}{\STy}{\STy_0})\in\valenv
  \end{array}
  }
\]
\end{center}
\end{figbox}
\caption{Operational Semantics for \protect{\Sdata}: Dataflow Semantics}
\label{fig:operational-semantics-Sdata}
\end{figure}

\begin{figure}
\begin{figbox}
\[  \parFlowstate{\ProcFlowstateSym}{\emptyFlowstate} \equiv \ProcFlowstateSym
     \Hskip
      \parFlowstate{\ProcFlowstateSym_1}{\ProcFlowstateSym_2}\equiv 
       \parFlowstate{\ProcFlowstateSym_2}{\ProcFlowstateSym_1}
 \] \[
   \parFlowstate{(\parFlowstate{\ProcFlowstateSym_1}{\ProcFlowstateSym_2})}{\ProcFlowstateSym_3}
  \equiv
   \parFlowstate{\ProcFlowstateSym_1}{(\parFlowstate{\ProcFlowstateSym_2}{\ProcFlowstateSym_3})}
\] \[  \seqFlowstate{\ProcFlowstateSym}{\emptyFlowstate} \equiv \ProcFlowstateSym
     \Hskip
      \seqFlowstate{\ProcFlowstateSym_1}{\ProcFlowstateSym_2}\equiv 
       \seqFlowstate{\ProcFlowstateSym_2}{\ProcFlowstateSym_1}
 \] \[
   \seqFlowstate{(\seqFlowstate{\ProcFlowstateSym_1}{\ProcFlowstateSym_2})}{\ProcFlowstateSym_3}
  \equiv
   \seqFlowstate{\ProcFlowstateSym_1}{(\seqFlowstate{\ProcFlowstateSym_2}{\ProcFlowstateSym_3})}
  \]
\[ 
\infer{
       \aredTyj{\tyenv}{\CausalitySetSym}{(\parFlowstate{\ProcFlowstateSym_1}{\ProcFlowstateSym_2})}
                              {\eventSym}{(\parFlowstate{\ProcFlowstateSym_1'}{\ProcFlowstateSym_2})}
    }{
       \aredTyj{\tyenv}{\CausalitySetSym}{\ProcFlowstateSym_1}{\eventSym}{\ProcFlowstateSym_1'}
    }
\]      
\[ 
\infer{
       \aredTyj{\tyenv}{\CausalitySetSym}{(\seqFlowstate{\ProcFlowstateSym_1}{\ProcFlowstateSym_2})}
                              {\eventSym}{(\seqFlowstate{\ProcFlowstateSym_1'}{\ProcFlowstateSym_2})}
    }{
       \aredTyj{\tyenv}{\CausalitySetSym}{\ProcFlowstateSym_1}{\eventSym}{\ProcFlowstateSym_1'}
    }
\]      
\[
\infer{
        \airedTyj{\tyenv}{\CausalitySetSym}
                   {\eventComp{\eventSym}{(\seq{\iterTypeSym},\iterType{t}{\numberTy{m}}{\numberTy{n}}),\seq{\guardTypeSym}}}
                   {\seqFlowstate{(\subst{\eventComp{\eventSym}{\seq{\iterTypeSym},\seq{\guardTypeSym}}}{\numberTy{m}}{t})}
                                          {\eventComp{\eventSym}{(\seq{\iterTypeSym},\iterType{t}{(\numberTy{m+1})}{\numberTy{n}}),\seq{\guardTypeSym}}}}
   }{
     m \leq n
  }
\]
\[
\infer{
        \airedTyj{\tyenv}{\CausalitySetSym}
                   {\eventComp{\eventSym}{(\seq{\iterTypeSym},\iterType{t}{\numberTy{m}}{\numberTy{n}}),\seq{\guardTypeSym}}}
                   {\eventComp{\eventSym}{\seq{\iterTypeSym},\seq{\guardTypeSym}}}
   }{
     m > n
  }
\]
\[
\infer{
        \airedTyj{\tyenv}{\CausalitySetSym}
                   {\eventComp{\eventSym}{\seq{\iterTypeSym},(\seq{\guardTypeSym},\numberTy{m}\relop\numberTy{n})}}
                   {\eventComp{\eventSym}{\seq{\iterTypeSym},\seq{\guardTypeSym}}}
   }{
     m \relop n
  }
\]
\[
\infer{
        \airedTyj{\tyenv}{\CausalitySetSym}
                   {\eventComp{\eventSym}{\seq{\iterTypeSym},(\seq{\guardTypeSym},\numberTy{m}\relop\numberTy{n})}}
                   {\emptyFlowstate}
   }{
     \neg(m \relop n)
  }
\]
\[
\infer{
        \aredTyj{\tyenv}{\CausalitySetSym}
                   {\seqFlowstate{\eventSym}{\ProcFlowstateSym}}
                   {\eventSym}
                   {\ProcFlowstateSym}
   }{
  }
\]
\[
\infer{
      \aredTyj{\tyenv}{\CausalitySetSym}{\ProcFlowstateSym_1}{\eventSym}{\ProcFlowstateSym_2}
  }{
      \ProcFlowstateSym_1\equiv\ProcFlowstateSym_1'
      \Hskip
      \aredTyj{\tyenv}{\CausalitySetSym}
                 {\ProcFlowstateSym_1'}{\eventSym}{\ProcFlowstateSym_2'}
       \Hskip
       \ProcFlowstateSym_2\equiv\ProcFlowstateSym_2'
  }
\]
\end{figbox}
\caption{Type reduction rules}
\label{fig:type-reduction-rules}
\end{figure}

\begin{figure}
\begin{figbox}
\begin{center}
$(\val\parsym\Proc)\equiv \Proc$
\Hskip
$(\Proc_1\parsym\Proc_2)\equiv (\Proc_2\parsym\Proc_1)$

\Vskip
$(\Proc_1\parsym(\Proc_2\parsym\Proc_3))\equiv
    ((\Proc_1\parsym\Proc_2)\parsym\Proc_3)$
\[
\infer{
\eventComp{\eventSym}{\iterType{t}{\numberTy{m}}{\numberTy{n}}} \equiv
    (\subst{\eventSym}{\numberTy{m}}{t} \parsym \ldots \parsym
     \subst{\eventSym}{\numberTy{n}}{t} )
}{
  m\leq n
}
\]
%
%
\end{center}
\end{figbox}
\caption{Structural equivalence for processes}
\label{fig:structural-equivalence}
\end{figure}

\begin{figure}
\begin{figbox}
\begin{center}
\[
\infer[\text{\sc Red App}]{
  \Config{\val(\seq{\val})}{\Heap}
  \iredExp
  \Config{\subst{\Exp}{\seq{\val}}{\seq{x}}}{\Heap}
}{
  \val=(\lambda \seq{x}{:}\seq{\STy}{.}
      \lambdar{\ActorFlowstateSym_1}{\ActorFlowstateSym_2}
      \Exp)
}
\]
\[
\infer[\text{\sc Red Let}]{
  \Config{(\kw{let}\ x=\val\ \kw{in}\ \Exp)}{\Heap}
  \iredExp
  \Config{\subst{\Exp}{\val}{x}}{\Heap}
  }{
  }
\]
\[
\infer[\text{\sc Red Deref}]{
  \Config{\derefExp{l}}{\Heap}
  \iredExp
  \Config{\Heap(l)}{\Heap}
  }{
  }
\]
\[
\infer[\text{\sc Red Assign}]{
  \Config{\assignExp{l}{\val}}{\Heap}
  \iredExp
  \Config{\val}{\Heap'}
  }{
    \Heap'=\Heap[\heapBind{l}{\val}]
  }
\]
\[
 \infer[\text{\sc Red If True}]{
  \Config{(\kw{if}\ \kw{true}\ \kw{then}\ \Exp_1\ \kw{else}\ \Exp_2)}{\Heap}
  \iredExp
  \Config{\Exp_1}{\Heap}
  }{
  }
\] 
\[\infer[\text{\sc Red If False}]{
  \Config{(\kw{if}\ \kw{false}\ \kw{then}\ \Exp_1\ \kw{else}\ \Exp_2)}{\Heap}
  \iredExp
  \Config{\Exp_2}{\Heap}
  }{
  }
\]
\[
  \infer[\inflab{\sc Red Exp Cong}]{
    \Config{\ExpCtxt[\Exp_1]}{\Heap}
    \optRedProc{\valenv}{\eventSym}
    \Config{\ExpCtxt[\Exp_2]}{\Heap'}
  }{
    \Config{\Exp_1}{\Heap}
    \optRedProc{\valenv}{\eventSym}
    \Config{\Exp_2}{\Heap'}
  }
\]
\[
\infer[\inflab{\sc Red Proc Cong}]{
    \Config{\ProcCtxt[\Proc_1]}{\Heap}
    \optRedProc{\valenv}{\eventSym}
    \Config{\ProcCtxt[\Proc_2]}{\Heap'}
  }{
    \Config{\Proc_1}{\Heap}
    \optRedProc{\valenv}{\eventSym}
    \Config{\Proc_2}{\Heap'}
  }
\]
\end{center}
\end{figbox}
\caption{Operational Semantics for \protect{\Sdata}: Core Semantics}
\label{fig:operational-semantics-Sdata-core}
\end{figure}

Our basic result is that evaluation preserves types, in the sense that
a type may simulate the communications performed at the value level:
\begin{theorem}[Type Preservation]
If \procj{\tyenv}{\valenv}{\Proc_1}{\ProcFlowstateSym_1} and
\heapj{\tyenv}{\valenv}{\Heap_1}{\ProcFlowstateSym_1'}, and
$\Config{\Proc_1}{\Heap_1} \optRedProc{\valenv}{\eventSym}
   \Config{\Proc_2}{\Heap_2}$,
then \procj{\tyenv}{\valenv}{\Proc_2}{\ProcFlowstateSym_2} and
\heapj{\tyenv}{\valenv}{\Heap_2}{\ProcFlowstateSym_2'}, for some
$\ProcFlowstateSym_2$ and $\ProcFlowstateSym_2'$, where
\aredTyj{\tyenv}{}{\ProcFlowstateSym_1}
                           {\eventSym}
                           {\ProcFlowstateSym_2} and one of the
                           following holds:
\begin{enumerate}
\item Either $\mathit{Producer}(\tyenv,\eventSym)$ is true and
  $\ProcFlowstateSym_2'\equiv(\parFlowstate{\ProcFlowstateSym_1'}{\eventSym})$; or
\item $\mathit{Consumer}(\tyenv,\eventSym)$ is true and
  $\ProcFlowstateSym_1'\equiv(\parFlowstate{\ProcFlowstateSym_2'}{\eventCompl{\eventSym}})$.
\end{enumerate}
\end{theorem}

Our progress result reflects that computation is not deadlocked,
provided the initial heap is compatible with the remaining actor
computation, as reflected in the flowstate.  Note that this result is
for a single firing of the dataflow graph; for simplicity, we do not
consider the unrolling of the graph for another firing.
\begin{theorem}[Progress]
If \procj{\tyenv}{\valenv}{\Proc_1}{\ProcFlowstateSym_1} and
\heapj{\tyenv}{\valenv}{\Heap_1}{\ProcFlowstateSym_1'}, 
and \livefsj{\tyenv}{\ProcFlowstateSym_1}{\ProcFlowstateSym_1'}{\ProcFlowstateSym_2}{\ProcFlowstateSym_2'},
then $\Config{\Proc_1}{\Heap_1} \optRedProc{\valenv}{\eventSym}
   \Config{\Proc_2}{\Heap_2}$, for some $\Proc_2$, $\Heap_2$ and $\eventSym$.
\end{theorem}

Our dataflow language elides any compositional constructs for building
dataflow graphs ``bottom-up.''  Examples of operators that might be
used for such compositional constructs are provided elsewhere
\cite{dataflow:TBRL10:tr,dataflow:DY12:ecoop}.  
Central to that work is the abstraction of the communication structure
of a component dataflow graph, exposing causal dependencies in
communication channels to prevent deadlock during incremental
construction of a dataflow graph.
An interesting
direction for future research would be to consider how to combine the
parameterized dataflow considered in this report with those causalities.

%% file: related.tex
\section{Related Work}
\label{sect:related}

The notion of types that describe resource usage
largely come out of the realm of linear \cite{types:Wadler90,types:Cyclone02,types:MR02,types:DF01}
and affine 
\cite{types:TP11}
 type systems for statically checking the safe usage of limited
resources.  
The typical approach is to provide a
linear type system where we are guaranteed exactly one reference to a
resource.  
Two particularly significant lines of study in the ``linear types'' field have
been the approach of \emph{typestate} \cite{types:DF04}
and that of \emph{session types} \cite{types:DMD10}. 
\iftoggle{short}{
The current work uses the framework of sessional dataflow, that
combines dataflow with session types.
\emph{Usage types} \cite{types:Kob03}  have a similar motivation to sessional dataflow,
statically preventing the composition of concurrent components that
would produce deadlocks.  
The approach of session types \cite{types:DMD10} is commonly motivated by its support for safe
Web services.
Session types have been realized in both functional \cite{types:VAS04} and object-oriented
language 
\iftoggle{short}{
        \cite{types:CCDDG09} 
}{
        \cite{types:MDNS06,types:CDY07,types:CCDDG09} 
}
semantics, with both synchronous and asynchronous semantics.
Dyadic session types have also been generalized to multiparty \cite{types:HYC08} interactions,
where potentially more than two parties are involved in an interaction.  This approach
has further been generalized to a dynamically varying number of participants,
based on assigning roles \cite{types:DY11} to participants and describing generic protocols for
each participant role.  

Another related line of work is in synchronous languages for real-time and
embedded systems. The constraints on the synchronous languages preclude any need for
buffering, since all actors operate in lock step on the same
clock\footnote{Lustre and its descendants allow multiple clocks, used by
different components, but all clocks have a common base clock.  A true
multi-clock synchronous language is the Signal language, but consideration of
Signal is outside the scope of the current work.
}. The theory of $N$-synchronous Kahn networks \cite{synch:Cetal06}
relaxes the synchrony
restriction, allowing different actors to have their own clock rates, and
allowing buffering between actors to match their clock rates.  It is
therefore related to the approach of synchronous dataflow \cite{synch:LM87}.  While
$N$-synchronous Kahn networks uses the different clock speeds of the actors to
compute the amount of buffer space required, and to schedule the execution of
the actors, SDF is instead using the data rates of the actors on their input
and output channels to compute buffer sizes and perform scheduling. 

}{

 Typestate is a
concept that originated in the Hermes language of Strom and Yemini \cite{types:RES86}.  It
corresponds to an enrichment of the normal notion of a type, to include the
concept of types as states in a finite state machine.  F\"ahndrich and Deline
incorporated this idea into object oriented languages \cite{types:DF04} in a very natural way:
each object has a typestate, and the interface offered by an object, in the sense
of the methods that can currently be invoked on the object, are determined by
its current typestate.  Since typestate is updated imperatively, it is
important that aliasing of such objects be carefully controlled.  Aldrich et
al \cite{dataflow:SMAJ09} have demonstrated that a notion of \emph{permissions}, based on earlier
work on type-based capabilities, can be used to check the use of typestate in
existing non-toy software systems.  We have explicitly avoided introducing
these issues into the current report, but they are clearly relevant to
incorporating sessional dataflow into real programming languages.

The approach of session types \cite{types:DMD10} is commonly motivated by its support for safe
Web services.  In the simplest case, session types are used to mediate the
exchanges between two parties in a dyadic interaction.  Each session offers a
``shared channel'' (different from our use of the terminology), essentially an
service endpoint URL that a client connects to.  On connection, a new server
thread is forked and a private session channel is established between 
the client and this thread.  This channel has a behavioral type that is
essentially an abstract single-threaded process, that constrains the
communications between the parties.  These constraints include ensuring that
both parties agree on the type of data exchanged, when the type of the data
may vary depending on where the parties are in the protocol, and ensuring that
parties agree on the possible decision branches that may be reached at various
points in the protocol.  Typically the server offers a nondeterministic
external choice, and the client makes the selection of one of these choices.
Since only the client and the server share their private channel, the
execution is in fact deterministic.  

Session types have been realized in both functional \cite{types:VAS04} and object-oriented
language 
\iftoggle{short}{
        \cite{types:CCDDG09} 
}{
        \cite{types:MDNS06,types:CDY07,types:CCDDG09} 
}
semantics, with both synchronous and asynchronous semantics.  In the
latter asynchronous case, the session type ensures an upper bound on the total size
required of the communication buffers.  Functional languages incorporating
session types are typically based on intuitionistic linear logic, i.e., using
the framework of linear types to model much of the resource usage machinery
required for session types.  For example, a standard extension of session
types is to allow the server to delegate all or some of its processing to
another server, by sending the private channel it shares with the client to
another server.  It is critical that the delegating server not continue to
communicate on the private channel, i.e., that there be no dangling
references, and this is ensured by requiring that the channel be used
linearly.  In the functional setting, issues of variable aliasing are not so
critical, and  type systems focus on proper resource usage analysis.  Object
oriented languages incorporating session types replace the notion of method
invocation with sessions.  Rather than transmitting arguments from client to
server all at once at the point of method invocation, the arguments are
transmitted over time in the process of a protocol where the client interacts with the
server to ``negotiate'' on the nature of the functionality the client wishes
from the server.

Dyadic session types have also been generalized to multiparty \cite{types:HYC08} interactions,
where potentially more than two parties are involved in an interaction.  The
approach is to specify a global session protocol as an abstract parallel
program, and then extract the ``obligations'' of each participant in the
protocol as a single-threaded program in an automatic fashion.  This approach
has further been generalized to a dynamically varying number of participants,
based on assigning roles \cite{types:DY11} to participants and describing generic protocols for
each participant role.  

Although sessional dataflow might appear at first related to session types,
the connection is actually rather weak, because of the nature of the
interactions in dataflow.  Whereas session types provide a very fine-grained
specification of the interaction between parties in a concurrent interaction,
the main attraction of dataflow is that the parties are loosely coupled, and
this flexibility is an important part of the usefulness of session types for
parallelizing applications.  The closest this system comes to a session types
system is in the behavioral constraint on the behavior of an actor, in terms
of matching the specified input and output data rates on each firing specified
on an actor interface.  However this behaviorial specification only constrains
a single actor, and places no constraint on the behavior of its neighboring
actors (upstream or downstream).  Furthermore a session type
specifies, for each participant in an interaction, a very precise
single-threaded behavior, in terms of data exchanged on the private session
channels at each point in the execution.  In contrast, most of the behavioral
specification for an actor in sessional dataflow is concurrent, with causality
specified only
between consumption of input messages and production of output messages, and
between the starting and ending typestates in the case of cyclostatic
dataflow.


Another related line of work is in synchronous languages for real-time and
embedded systems.  Such languages assume a ``clock'' on all computations, with
variables representing potentially infinite streams of values, indexed by clock
ticks.  Here the most relevant  example for sessional dataflow is that of
Lustre \cite{synch:Lustre}, a language that 
is a dataflow language in the tradition of Lucid \cite{synch:Lucid}, and is a synchronous
language in the sense of the synchronous languages such as Esterel \cite{synch:BS91}, but which
we cannot call a synchronous dataflow language for fear of confusing the
reader.  The constraints on the synchronous languages preclude any need for
buffering, since all actors operate in lock step on the same
clock\footnote{Lustre and its descendants allow multiple clocks, used by
different components, but all clocks have a common base clock.  A true
multi-clock synchronous language is the Signal language, but consideration of
Signal is outside the scope of the current work.
}.  The theory of these ``synchronous,'' ``dataflow'' networks has been
described in terms of synchronous Kahn networks \cite{synch:CP96}, which have the property that
no buffering is required at all between actors, since all execution is
synchronous and
governed by a common clock.   This is clearly a very
strong restriction, albeit one that facilitates compilation of programs to
hardware circuits.  The theory of $N$-synchronous Kahn networks \cite{synch:Cetal06} relaxes this
restriction, allowing different actors to have their own clock rates, and
allowing buffering between actors to match their clock rates.  It is
therefore very much related to the approach of synchronous dataflow \cite{synch:LM87}.  While
$N$-synchronous Kahn networks uses the different clock speeds of the actors to
compute the amount of buffer space required, and to schedule the execution of
the actors, SDF is instead using the data rates of the actors on their input
and output channels to compute buffer sizes and perform scheduling.  Since
ultimately our interest is in compiling dataflow programs to hardware, this is
an intriguing connection that we intend to explore in future work. 

\emph{Usage types} \cite{types:Kob03}  have a similar motivation to sessional dataflow,
statically preventing the composition of concurrent components that
would produce deadlocks.  Sessional dataflow chooses a specialized
programming model, dataflow, and tracks exact causalities, whereas
usage types choose a general message-passing programming model, and
track potential causalities using a form of logical time.  A
by-product of usage types is that they can be used to compute upper
bounds on the amount of resource usage (semantic reductions) used for
execution.  This is related to the notion of \emph{resource bound
  certification} \cite{types:CK00}, that statically computes upper bounds on the amount
of time taken to execute a piece of code.  The timing analysis for
dataflow programs in this article is similar, and in fact deliberately
simpler, since the timing analysis is not the main point of the
paper.  One consequence of the simplified nature of our timing
analysis is that we assume that all communications for an actor occur
at the end of its execution, so that any ``clients'' of its outputs
cannot be scheduled until after that actor has completed.  It would
clearly be advantageous to have a timing analysis that provided more
fine-grained information, e.g., putting bounds on when
individual communication events are estimated to occur.  We have
developed such an analysis, and hope to report on it in a future article.

}

%% file: concl.tex
\section{Conclusions}
\label{sect:concl}

We have described a type and effect system for a dataflow language,
that allows the firing rates of actors to be parameterized in their
description, while allowing modular analysis of actor bodies for their
firing behavior.  An obvious direction for future research is to
consider the combination of this with the compositional sessional
dataflow system described in \cite{dataflow:DY12:ecoop}.  The main
challenge here is the assumption, in the system described in this
paper, that there are no causal dependencies between communications
in different iterations of a loop.